\documentclass{aa} 

\usepackage{txfonts}
\usepackage{graphicx}   
\usepackage{amsmath}    
\usepackage[utf8]{inputenc} 
\usepackage{array} 
\usepackage{multicol,rotating} 
\usepackage{longtable} 
\usepackage[section]{placeins}
\usepackage{lscape}
\usepackage{multirow} 

\begin{document}

   \title{Phobos photometric properties from Mars Express HRSC observations}

   \author{S. Fornasier
          \inst{1,2}
          \and
          A. Wargnier\inst{1} \and
          P.H. Hasselmann\inst{1} \and
          D. Tirsch\inst{3} \and
          K.-D. Matz\inst{3} \and
          A. Doressoundiram\inst{1}\and
          T. Gautier\inst{4}\and
          M.A. Barucci\inst{1}  
           }

   \institute{LESIA, Universit\'e Paris Cit\'e, Observatoire de Paris, Universit\'e PSL, CNRS, Sorbonne Universit\'e, 5 place Jules Janssen, 92195 Meudon, France
              \email{sonia.fornasier@obspm.fr}
         \and
           Institut Universitaire de France (IUF), 1 rue Descartes, 75231 PARIS CEDEX 05  
        \and  
         DLR-Institute of Planetary Research, Rutherfordstrasse 2, 12489 Berlin Germany
        \and
        LATMOS, 11 Boulevard d'Alembert, 78280 Guyancourt, France
             }

   \date{Received on January 2024}

 
  \abstract
{}
   {{This study aims to analyze Phobos' photometric properties using Mars Express mission observations to support the Martian Moons eXploration mission (MMX) devoted to the investigation of the Martian system and to the return of Phobos samples.}}
{We analyzed resolved images of Phobos acquired between 2004 and 2022 by the High Resolution Stereo Camera (HRSC) on board the Mars Express spacecraft at a resolution ranging from $\sim$ 30 m~px$^{-1}$ to 330 m~px$^{-1}$. We used data acquired with the blue, green, red, and IR filters of HRSC and the panchromatic data of the Super Resolution Channel (SRC). The SRC data are unique because they cover small phase angles (0.2-10$^o$), permitting the investigation of the Phobos opposition effect. We simulated illumination and geometric conditions for the different observations using the Marx Express and the camera spice kernels provided by the HRSC team. We performed photometric analysis using the Hapke model for both integrated and disk-resolved data.}
   {The Phobos phase function is characterized by a strong opposition effect due to shadow hiding, with an amplitude and a half-width of the opposition surge of 2.28$\pm$0.03 and 0.0573$\pm$0.0001, respectively. Overall, the surface of Phobos is dark, with a geometric albedo of 6.8 \% in the green filter and backscattering. Its single-scattering albedo (SSA) value (7.2\% in the green filter) is much higher than what has been found for primitive asteroids and cometary nuclei and is close to the values reported in the literature for Ceres. We also found a surface porosity of 87\%, indicating the presence of a thick dust mantle or of fractal aggregates on the top surface. The SSA maps revealed high reflectance variability, with the blue unit area in the northeast Stickney rim being up to 65\% brighter than average, while the Stickney floor is among the darkest regions, with reflectance 10 to 20\% lower than average. Photometric modeling of the regions of interest selected in the red and blue units indicates that red unit terrains have a stronger opposition effect and a smaller SSA value than the blue ones, but they have similar porosity and backscattering properties.}
   {The HRSC data provide a unique investigation of the Phobos phase function and opposition surge, which is valuable information for the MMX observational planning. The Phobos opposition surge, surface porosity, phase integral, and spectral slope are very similar to the values observed for the comet 67P and for Jupiter family comets in general. Based on these similarities, we formulate a hypothesis that the Mars satellites might be the results of a binary or bilobated comet captured by Mars.}


   \maketitle
%

\section{Introduction}


\begin{figure*}
    \centering
    \includegraphics[width=0.97\textwidth,angle=0]{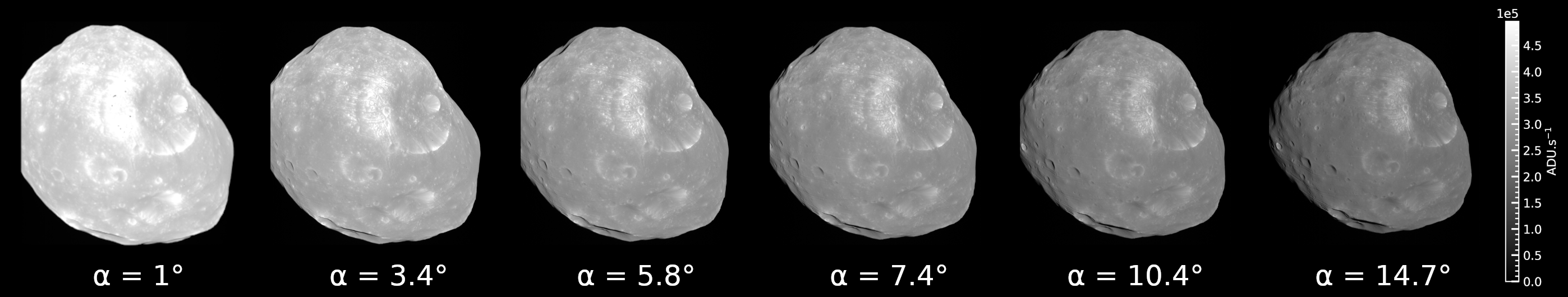}
    \includegraphics[width=0.97\textwidth,angle=0]{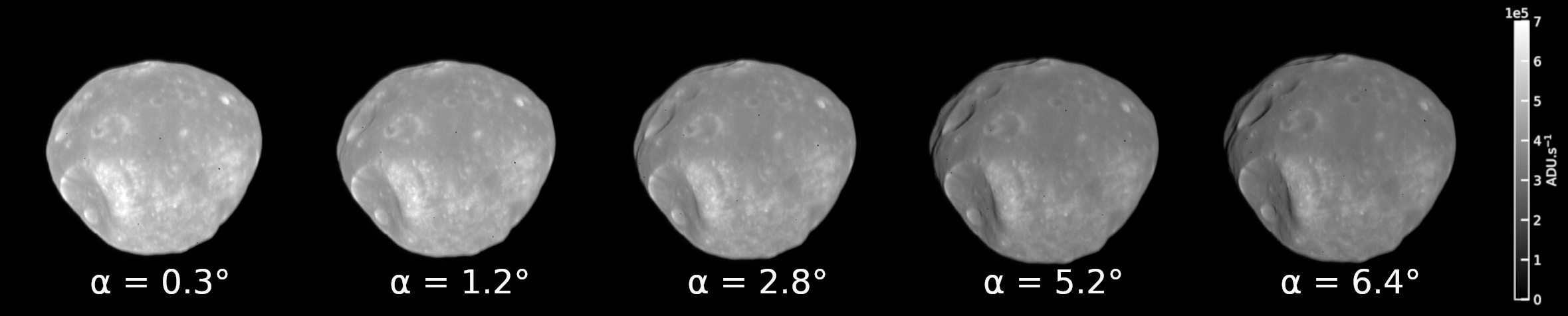}
    \caption{Images showing the flux increasing with a decreasing phase angle from images acquired during the orbit K076 (top panel) and M247 (bottom panel) with SRC. The flux is in DU s$^{-1}$; no other normalization was applied here (the heliocentric distance was the same for a given orbit).}
    \label{SRCimages}
\end{figure*}

Phobos and Deimos, the moons of Mars, are quite peculiar in relation to other moons in the Solar System. They are very small compared to Mars ($\sim$ 15-25 km in size), and they have heavily cratered surfaces and a relatively low density. 
Phobos, in particular, displays a complex geomorphology, characterized by some peculiar 1-km long grooves. Several of its craters show various states of degradation. Notably, the Stickney crater has a diameter of up to 9.4 km \citep{Thomas_1999}. Phobos is covered by a layer of regolith 5 to 100 m thick \citep{Basilevsky_2014}. Its surface is dark, as it is for Deimos, with a low geometric albedo ($ \sim$ 7\%) in the V filter \citep{Zellner_1974}. Its bulk density is about half of Mars (1.85$\pm$0.07 g cm$^{-3}$, \cite{Pieters_2014}).

The surface composition is heterogeneous, and two distinct units, blue and red, have been identified since the first spatially resolved spectra acquired by the Phobos 2 mission \citep{Murchie_1996, Bibring_1992}. The blue unit shows a moderate slope spectrum and higher reflectance, while the red unit is darker and has a steeper slope, especially in the near-infrared region. The red unit dominates the majority of the surface of Phobos, while the blue one is mostly located around and partially inside the Stickney crater (it was initially considered a fresh ejecta deposit). Further high-resolution images acquired with the HIRISE instrument on board the Mars Reconnaissance Orbiter (MRO) indicated a more complex mixing in composition, with spots of redder material within the blue unit areas \citep{Basilevsky_2014}. \\

The origin of the satellites of Mars is debated, and two theories have been proposed. One theory is that Phobos and Deimos are asteroids captured by Mars (Hansen et al., 2018), a hypothesis reinforced by the spectral similarity between the two satellites and primitive D-type asteroids, which are found mainly in the outer asteroid belt and among Jupiter’s Trojans. However, dynamical simulations of the possible gravitational capture scenario fail to explain their quasi-equatorial orbits (Burns 1992). The spectral analogy with D-type asteroids suggests that the Martian satellites, if they are indeed captured asteroids, may be rich in organic material, carbon and hydrous silicates, and may possibly contain water ice in their interiors \citep{Rivkin_2002}. The second theory claims that the satellites formed in situ from a debris disk produced by the collision of a large object with Mars \citep{Craddock_2011, Rosenblatt_2012}. This scenario explains the orbits of these moons, but it does not explain their carbonaceous-like compositions \citep{Nakamura_2021}.

The spectra of Phobos are usually featureless, except for a 0.65 $\mu$m absorption feature observed in the red unit and a sharp 2.8 $\mu$m feature observed in both units \citep{Fraeman_2014}. The red unit feature centered at 0.65 $\mu$m was observed by a ground-based telescope \citep{Murchie_2015} and from different mission data \citep{Simonelli_1998, Fraeman_2014, Pajola_2013}, and it has been attributed to nontronite, a desiccated iron bearing phyllosilicate \citep{Murchie_2008}, or to space weathered anhydrous silicates \citep{Clark_2012}. The 2.8 $\mu$m feature is stronger in the red unit spectra and attributed to structural hydroxyl embedded in mineral lattices \citep{Rivkin_2002, Fraeman_2014}.


The debate on the origin of the Martian satellites should soon be resolved thanks to samples that will be returned by the JAXA Martian Moons eXploration (MMX) mission, which is devoted to the exploration of the Mars system. The MMX mission will be launched in 2026, be inserted into Mars orbit in 2027, and investigate the Martian system for three years, focusing mainly on Phobos, the key target of the mission. The main goals of MMX are to return samples of Phobos and, through both a detailed in situ investigation of the Mars satellites and further laboratory studies of Phobos samples on Earth, to clarify the origin of the Mars satellites and the process of planet formation in the Solar System.

In this work, we perform the very first analysis of Phobos photometric properties using data from the High Resolution Stereo Camera (HRSC) on board the Mars Express spacecraft. This work will support the MMX mission planners in the optimization of the trajectories and science operations, especially in defining the exposure time needed to achieve the expected signal-to-noise ratio of a given Phobos observation.\\
Photometric investigations of Phobos have been carried out in the past by \cite{Klaasen_1979}, \cite{Pang_1983}, \cite{Avanesov_1991}, \cite{Simonelli_1998}, \cite{Cantor_1999}, and \cite{Thomas_1999} and more recently by  \cite{Pajola_2012} and \cite{Fraeman_2012}. The HRSC data offer a unique opportunity to study in detail the Phobos opposition surge, that is, the non-linear increase of the reflectance for small phase angle values.

\section{Observations and data reduction}

\begin{figure}
    \centering
    \includegraphics[width=0.49\textwidth,angle=0]{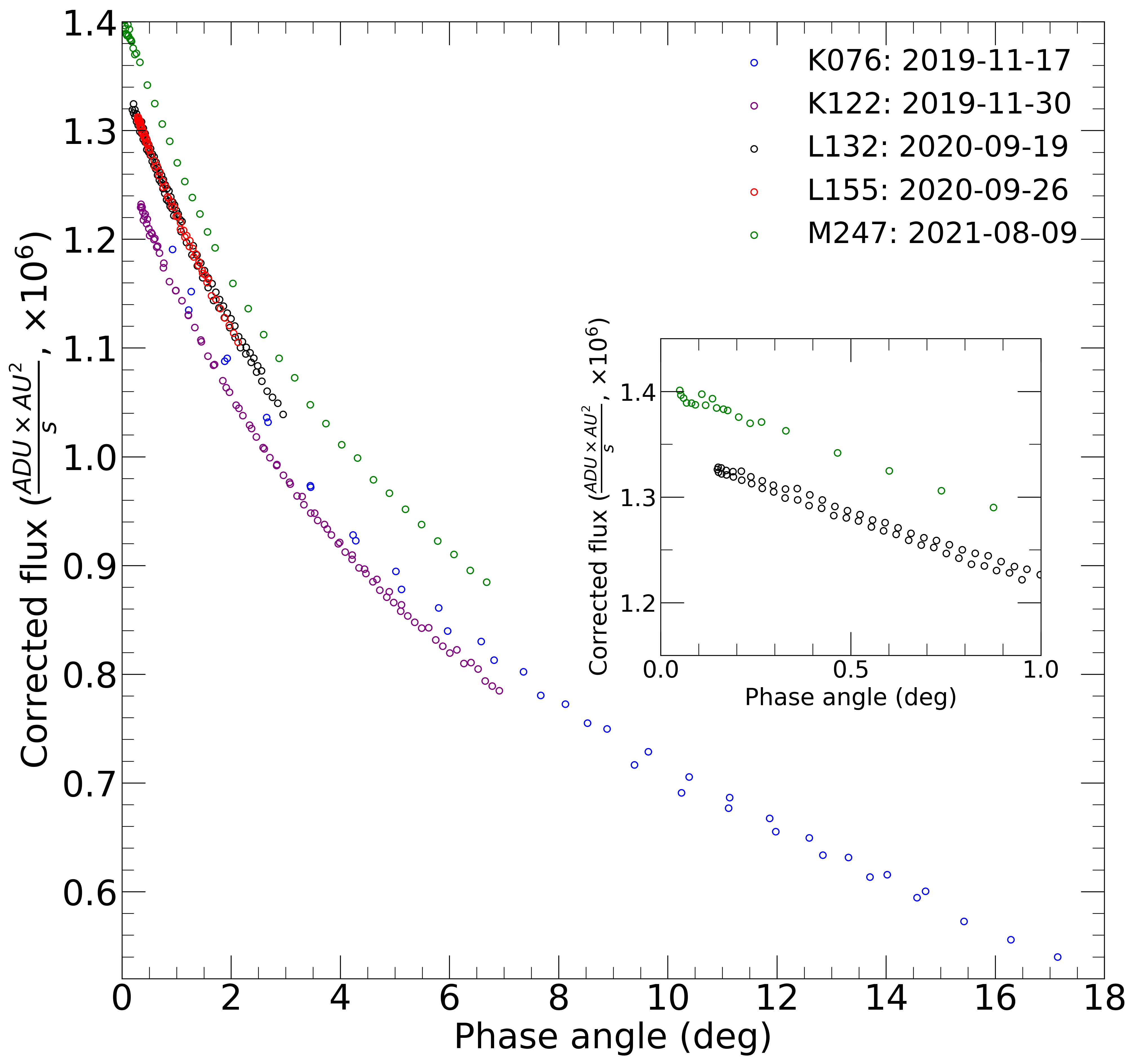}
    \caption{Observations of Phobos opposition effect from five individual SRC observations. Data have been normalized by the exposure time and corrected for heliocentric distance to the square. The insert shows very small phase angle data acquired for orbits L132 and M247. Data acquired at a phase angle lower than the angular size of the Sun seen from Phobos distance were discarded in the modeling.}
    \label{fig:SRC}
\end{figure}

\begin{figure}
    \centering
    \includegraphics[width=0.42\textwidth,angle=0]{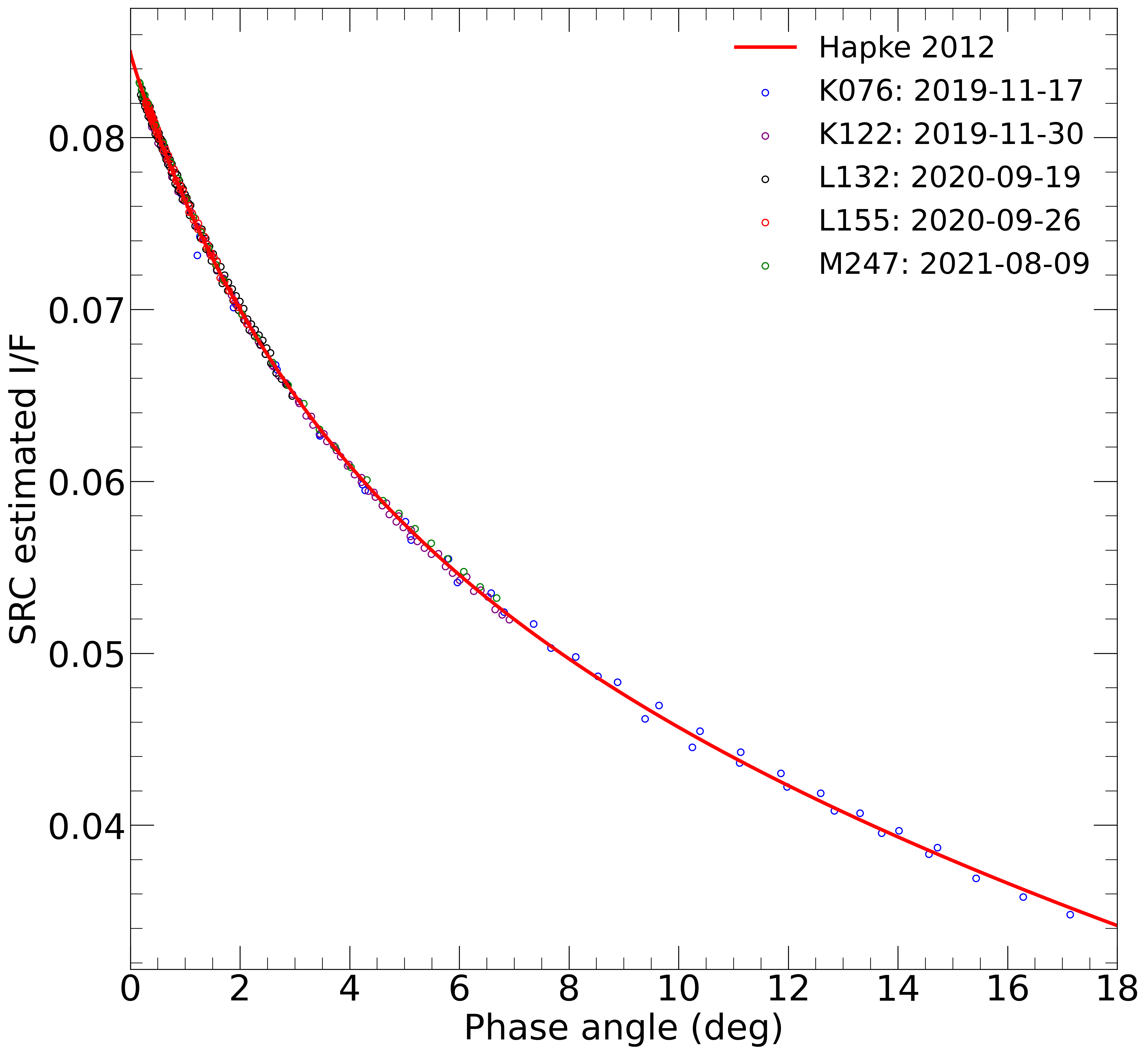}
    \includegraphics[width=0.42\textwidth,angle=0]{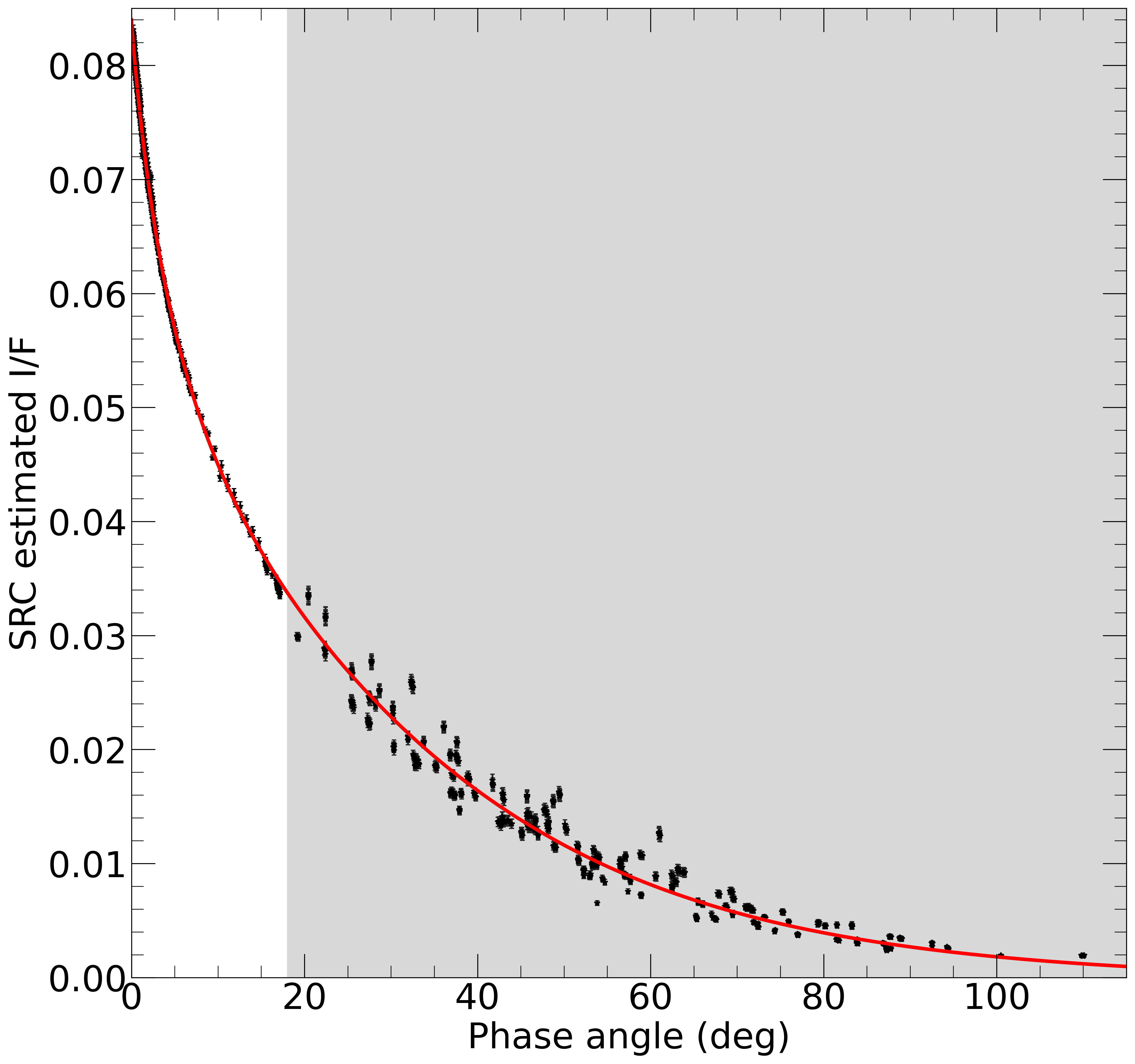}
    \caption{Integrated phase function of Phobos from SRC observations. Top: Opposition effect from integrated photometry from the five SRC orbits. The data presented in Fig.~\ref{fig:SRC} have been normalized, and the radiance factor was estimated from the measurement with the green filters at $\alpha$ = 5$^o$. Bottom: SRC phase function (black points) superposed on the Hapke global fit model. For completeness, some SRC data acquired during 2019-2020 at a higher phase angle are shown on the gray background, but these data were not used in the analysis.}
    \label{fig:SRC2}
\end{figure}

We investigated Phobos' photometric properties using the Mars Express HRSC data. The Mars Express mission was launched in 2003 by the European Space Agency, and it is principally devoted to the study of the Martian surface and atmosphere, especially to the observation of traces of water on the planet's surface. \\
The HRSC camera operates as a push-broom scanning instrument with an integrated stereo capability. It is used to observe the Martian surface and atmosphere as well as the planet’s two moons, Phobos and Deimos. This advanced camera system comprises nine CCD line detectors that capture highly detailed, simultaneous  high-resolution stereo image swaths, and it has the ability to incorporate multiple colors and phase angles \citep{Neukum_2004, Jaumann_2007}. Among the nine line sensors, four are equipped with spectral filters, enabling the acquisition of images in red, green, blue, and infrared wavelengths. The remaining five panchromatic sensors (nadir, stereo 1 + 2, photometry 1 + 2) serve to determine the photometric properties of the surfaces of the celestial bodies and enable the creation of digital terrain models (DTMs) \citep{Neukum_2004, Gwinner_2009}, whereas DTMs are only produced from observations of the Martian surface. The HRSC data calibration involves radiometric calibration for all acquired images and geometric correction for Mars surface observations. \\
The additional super resolution channel (SRC) of HRSC operates in panchromatic mode, utilizing a single CCD detector to acquire images with a significantly higher pixel resolution compared to the other color channels of the camera. The SRC captures images with a pixel size of 2.4 meters at an altitude of 250 kilometers (Jaumann et al., 2007) and was initially intended to allow for the detailed examination of geological features, surface textures, and small-scale structures on Mars. With higher resolution images available from other Mars orbiter missions, the main focus of the SRC turned to the observation of the Martian moons Phobos and Deimos as well as astrometric observations.
A typical calibration process for SRC could not be realized due to time and budget constraints during the mission preparation phase. In addition to the usual functional tests during the development and implementation process of the electronics, the only radiometric calibration measurements have been flat-field determinations by means of an integrating sphere and dark current measurements (DSNU). The calibration process of the SRC image data consists of four steps: (1) calculation of the average of the dark current pixels at the border of the CCD to get the dark signal uniformity (DSU) level, (2) subtraction of the product DSU$\times$DSNU from the image, (3) division of the image by the flat field, and (4) the flagging of hot pixels. This process permits the elimination of hot pixels generated on the CCD sensor by solar flares during the Mars Express cruise phase \citep{Jaumann_2007}.\\
Phobos photometry observations are acquired whenever the observation opportunities meet favorable requirements in terms of spatial resolution and small phase angle coverage, that is, when Phobos is observable at relatively small phase angles ($<$10$^{o}$) and when the Mars Express-target distance is less than 10,000 km. Depending on the distance to the moon, either SRC observations with inertial pointing (at higher distances) or HRSC color observations (blue, green, red, and infrared filters) with a spacecraft slew pointing mode (at lower distances $\sim$ 6000 km) are planned. For very small phase angles ($<$1$^{o}$), SRC observations in spot tracking mode are used, where the spacecraft keeps the moon constantly in the center of the field of view. The resulting pixel resolutions range between 2.4 meters and 400 meters, depending on the observation distance and camera system used (i.e., HRSC or SRC).

From the ESA Planetary Science Archive, we first selected images acquired with the absolutely calibrated  blue, green, red, and IR HRSC  filters. These data cover mostly the 10-100$^o$ phase range, with very few observations at small phase angles. \\
Therefore, to study the opposition effect, we also analyzed the SRC observations.

\begin{figure*}
    \centering
    \includegraphics[width=0.7\textwidth,angle=0]{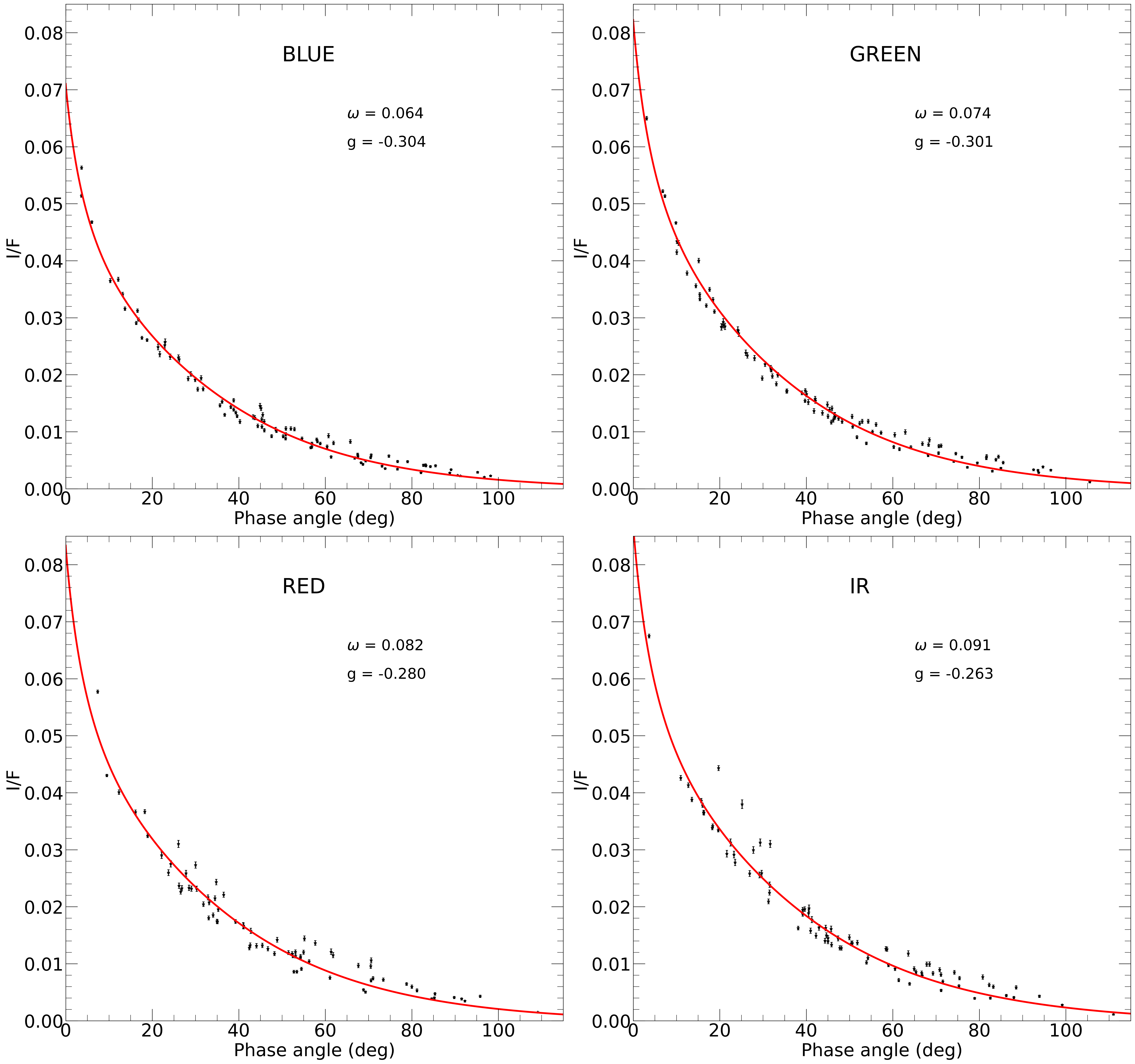}
    \caption{Phobos phase curve for the HRSC observations (black points) acquired with the blue, green, red, and IR filters. The red line represents the Hapke global fit model.}
    \label{HRSC_allintegrated}
\end{figure*}

\subsection{HRSC data analysis}

For the HRSC data in the absolutely calibrated blue, green, red, and IR filters, the radiance factor (also named $I/F$) was simply determined using the correction factors provided in the header of each image, which consider the absolute calibration values for each filter and the varying heliocentric distance.  The integrated flux was evaluated from aperture photometry divided by the projected Phobos surface, that is, considering the illuminated pixels and those in shadows derived for a given observation by the simulated images. To mimic the geometric conditions, namely, incidence ($i$), emission ($e$), phase angle ($\alpha$), longitude ($lon$), and latitude ($lat$), for each pixel, we generated simulated images using the latest version of the Mars Express mission and instrument spice kernels (including the latest shape, size and rotational constants\footnote{\url{https://naif.jpl.nasa.gov/pub/naif/generic_kernels/pck/pck00011_n0066.tpc}}). For the shape model, we utilized the stereophotogrammetric shape model\footnote{$PHOBOS\_K275\_DLR\_V02.BDS$} generated from Mars Express HRSC images \citep{Willner_2014}. 
For the simulation, we used the UTC mean time, that is, the time corresponding to the start time of a given acquisition plus half of the exposure time. \\
However, given the non-optimal NAIF/SPICE kernels, the generation of synthetic images was hindered. In fact, for several observations the simulated images were not in the nominal field of view of the camera, and non-systematic shifts in the field of view were needed. To improve the simulations, we adopted some field of view corrections files provided by the HRSC team that take into account the non-linearity of the pixel positions, but even with these corrections, some manual vertical and/or horizontal adjustments were still needed for a number of images. \\
In addition to this problem, we also discarded some simulations because their quality was not satisfactory. In fact, images were acquired in push-broom mode, and in some cases, the simulated images looked elongated and/or did not match completely with the original Phobos shape observed in a given image (Fig.~\ref{shape_nogood}). \\

We also had to apply image registration to improve the pixel matching between real and synthetic images. Given that HRSC image acquisitions are based on the push-broom mode, the instrument induces occasional distortions to the imaged cross section of the body. Moreover, the synthetic images were hindered by the NAIF/SPICE kernel's imprecision, which prevented the synthetic imaging from actually accounting for all distortion effects seen in the original images. In a past analysis performed by our team on space mission data, typical image co-registration of disk-resolved  images were performed, including corrections in offset, bi-dimensional rotation, and projective transformation \citep{Fornasier_2015, Feller_2016, Hasselmann_2017}. However, the mitigation of HRSC distortions must be dealt with by using a different technique. We therefore used the latest implementation of the TV-L1 dense optical flow technique \citep{Perez_2013}.
It computes the full flow field of displacement vectors with respect to changing features for pairs of images obtained under limited context alteration, that is, an entire body appearing in two images conserves most of the features in both frames. This estimated flow vector field transforms the synthetic images into similar pixel distributions of the original images, thus setting a correspondence between every pixel in the synthetic image with those in the original image. In our implementation, we fed the TV-L1 optical flow algorithm, available in the Python scikit-image package \citep{vanderWalt_2014}, synthetic images where Phobos' surface brightness was estimated using the Lommel-Seeliger law. The synthetic images were compared to original HRSC images that had been signal normalized in order to obtain the flow vector fields. 
Then, the transformed synthetic images were visually inspected in order to tune up the free parameters in the algorithm. 
To estimate the registration quality, we sorted the Universal Quality Image Index (UQI; \cite{Wang_2002}), which calculates the amount of similarity between two images. The UQI has been reported to perform better against noise as well as blurriness. When
compared to visual inspection, we set $UQI > 0.5$ as a satisfactory condition for similarity.
The final list of images on which we based our analysis is shown in Tables B.1, B.2, B.3, and B.4.

\subsection{Phobos opposition surge: SRC data analysis}

The SRC observations cover the Phobos opposition surge during six orbits from 2019 to 2022 (orbit numbers: K076, K122, K623, L132, L155, and M247). However, we ultimately did not use the K623 observations because they are partially contaminated by Mars straylight. In fact, during this observational session, Mars Express was on the dayside of Mars, in contrast to its position when the data were acquired during the other orbits. \\
These observations were optimized to investigate the opposition surge, covering the phase angle range 0.98-17.10$^o$ for K076, 0.4-7.0$^o$ for K122, 0.24-2.96$^o$ for L132, and 0.10-6.70$^o$ for M247, with a number of individual observations per orbit between 40 (K076) and 100 (L132). The final list of SRC images used to investigate the opposition surge is shown in Table B.5. \\ 
Examples of different phase angle images acquired in orbits K076 and M176 are presented in Fig.~\ref{SRCimages}, which clearly displays the increase of brightness as the phase angle decreases. 

Due to a lack of absolute calibrations, we trust only in relative photometry for SRC images. Data were corrected applying the following procedure:\\
a) We first divided each image by the exposure time because the data are provided in digital units and not normalized by the exposure time. b) We corrected the data by the heliocentric distance ($dist_{\odot}$) to the square, similar to what is done when generating the radiance factor, in order to consider the varying incoming solar flux. c) We computed the disk-integrated photometry for each observation simply as the integral of the flux from Phobos, evaluated using aperture photometry, divided by the object projected surface. The projected surface was estimated from simulated images generated using the latest Mars Express mission and instrument spice kernels available on the ESA planetary science archive and using the highest resolution Phobos shape model \citep{Ernst_2023}. \\
In contrast to the HRSC calibrated filter, the simulated images of the SRC framing camera match pretty well with the shape of the original ones. The co-registration was also quite straightforward, as only a simple translation of a few pixels was needed to match the real images. 

Some of the observations cover extremely small phase angles that are smaller than the angular size of the Sun as seen from Phobos:
\begin{equation}
\gamma_{sun} = \arcsin \frac{R_{\odot}}{dist_{\odot}},
\end{equation}
where R$_{\odot}$ is the radius of the Sun and dist$_{\odot}$ is the heliocentric distance of Phobos at a given observation. The angular size of the Sun is 0.16$^o$ for orbit M247 (dist$_{\odot}$ = 1.66 au) and 0.19$^o$ for orbit L132 (dist$_{\odot}$ = 1.39 au). For these two orbits, which cover the smallest phase angle, we considered in the photometric analysis only the observations acquired at $\alpha > \gamma_{sun}$ to avoid an underestimation of the opposition effect. In fact, it is well known that the brightness values flatten for $\alpha < \gamma_{sun}$ \citep{deau_2012}.

The integrated flux, in arbitrary units, is presented in Fig~\ref{fig:SRC}. All the data display a similar phase function behavior. In fact they are superposed after the normalization process described in the following (Fig.~\ref{fig:SRC2}). As the data are not absolutely calibrated, we attempted to estimate the I/F SRC data as follows, even if we stress that the I/F shown in Fig.~\ref{fig:SRC2} is not an absolutely calibrated value.
We first normalized the data in count since the observations during the different orbits were not at the same flux intensity. For this, we interpolated each orbit's data at a finer phase angle resolution, and then we normalized all the observations at $\alpha$=2$^{o}$. Finally, we multiplied the SRC data by a factor in order to match the calibrated reflectance values of the green filter at $\alpha$=5$^{o}$. This normalization factor is of course not perfect because the observations were acquired at different spectral ranges. In fact the SRC spectral range is larger than the green one. Moreover, the observed surface of Phobos is not the same. We point out that we utilized the SRC data to derive the width and amplitude of the opposition surge and not to derive Phobos' absolute reflectance.  
The estimated SRC I/F is presented in Fig.~\ref{fig:SRC2}.





\section{Disk-averaged photometry} 

\begin{table*}
\centering
\caption{Hapke parameters obtained from disk-averaged photometry.} 
\label{tab1}
{\footnotesize
\begin{tabular}{|cccccccl|}
  \hline
Camera & filter & $\omega$ & g & $B_0$ & h & $\bar{\theta}$ (deg) & geom. albedo \\
  \hline
SRC  & pan & -- & -0.265 & 2.283 $\pm$ 0.029 & 0.05728 $\pm$ 0.0005 & 24 &   -- \\ \hline
HRSC &  Blue           & 0.064 $\pm$ 0.001 & -0.304 $\pm$ 0.007 & 2.283 & 0.05728 & 24 &  0.0714\\
HRSC &  Green           & 0.074 $\pm$ 0.002 & -0.301 $\pm$ 0.007 & 2.283 & 0.05728 & 24 &  0.0816\\
HRSC &  Red           & 0.082 $\pm$ 0.003 & -0.280 $\pm$ 0.015 & 2.283 & 0.05728 & 24 &  0.0835\\
HRSC &  IR           & 0.091 $\pm$ 0.003 & -0.263 $\pm$ 0.014 & 2.283 & 0.05728 & 24 &  0.0877\\
  \hline
  \hline
Literature &&&&&&& Reference \\ \hline
Phobos & clear  &   0.070  & -0.13 & 4$^{+6}_{-1}$ & 0.055$\pm$0.025 & 22$\pm$2 & \cite{Simonelli_1998} \\    
Average C-ast   &v-band  & 0.037  & -0.47 & 1.03& 0.025 & 20 & \cite{Helfenstein_1989}\\
Mathilde  & v-band & 0.035$\pm$0.006  & -0.25$\pm$0.04 & 3.18$\pm$1.0 & 0.074$\pm$0.003 & 19$\pm$5 & \cite{Clark_1999}\\
Ceres$^{u}$     &v-band & 0.070  & -0.40 & 1.6 & 0.06  & 44 & \cite{Helfenstein_1989},  \\
                &       &      &       &      &      &    & \cite{Li_2006} \\ 
Ceres$^{r}$     & v-band &  0.116  & -0.22 & 1.6 & 0.054 & 25 & \cite{schroder_2018} \\
Ceres$^{r}$     & v-band &  0.104 & -0.31 & 1.6 & 0.06 & 18.7 & \cite{Li_2019}\\
Themis          &v-band &0.048  & -0.40  & 1.6 & 0.060 &  5 & \cite{Bowell_1989}\\
Ryugu           & v-band &  0.044  & -0.39  &  0.98 &  0.0.75 & 28 & \cite{Tatsumi_2020}\\
Bennu           & v-band  & 0.043  & -0.30  &  -- & -- & 14 & \cite{Golish_2021} \\      
67P$^{u}$       &  v-band   & 0.037  &-0.42  & 1.95& 0.023 & 15 & \cite{Fornasier_2015}\\ 
67P$^{r}$       &  red-band   & 0.034 & -0.42 & 2.25 & 0.061 & 28 & \cite{Fornasier_2015}\\ 
Moon            &  v-band &0.21 & -0.18 & 2.01 & 0.07 & 20 & \cite{Helfenstein_1987} \\
Mars (plains)$^{a}$   & r-band & 0.48 & -0.136 & 1.0 &  0.09 & 0 & \cite{Johnson_2006b} \\
Mars (dark rock)$^{b}$ & r-band & 0.53 & 0.229 & 0.77 &  0.5 & 11 & \cite{Johnson_2006a} \\
Mars (soil)$^{c}$ & r-band & 0.72 & -0.167 & 1.0 &  0.213 & 2 & \cite{Johnson_2006a} \\
\hline
\end{tabular}
}
\tablefoot{The SRC data were used to constrain the opposition surge of Phobos, and the derived $B_0$ and $h$ parameters were used as fixed values for the modeling of the four HRSC filters. We fixed $\bar{\theta}$ to the value obtained from the modeling of disk-resolved observations acquired with the green filter. Photometric parameters for other low-albedo small bodies, the Moon, and Mars are shown for comparison. For Mars, we report a few examples of photometric parameters at 743 nm from the Spirit and Opportunity rovers observations (we note that the $B_0$ parameter was underconstrained by data; see \cite{Johnson_2006b} and \cite{Johnson_2006a} for more details). Notations are as follows: $^{u}$ indicates results from disk-averaged photometry; $^{r}$ designates results from resolved photometry; $^{a}$ notes plains NW of Endurance (Table 3b in \cite{Johnson_2006b}) ; $^{b}$ indicates dark rock in Paso Robles 2 (Table 8d in \cite{Johnson_2006a}); $^{c}$ notes soil in Paso Robles 1 (Table 7c in \cite{Johnson_2006a}). }
\end{table*}


The disk-averaged photometry represents the global analysis of Phobos surface photometric properties. 
To model the Phobos photometry, we used the Hapke formalism following the same methodology and equations presented in \cite{Fornasier_2015}. For the global photometry, we used the disk-integrated Hapke expression \citep{Hapke_1993} with a single-term Henyey-Greenstein function {and neglecting the coherent-backscattering mechanism (CBOE), which is expected to have a low contribution for dark surfaces \citep{Shevchenko_2012}.} This model has five-parameters: the single-scattering albedo (SSA) $w_{\lambda}$, the asymmetry factor $g_{\lambda}$, the average roughness angle $\bar{\theta}$,
and two parameters defining the amplitude $B_{0}$ and the width $h$ of the shadow-hiding opposition surge. \\
Since observations at small phase angles are available only with SRC, we used the estimated SRC radiance factor on the combined observations from five orbits to determine the Phobos opposition surge parameters. As the Hapke parameters are not easy to disentangle and the roughness parameter may be determined only with observation at $\alpha$ $>$ 60$^0$, we fixed it at 24$^o$, the value determined from the modeling of the disk-resolved images obtained with the green filter (see Section 4).
We used the Levenberg-Marquardt algorithm { \citep{More_1978}} to find the best fit between the Hapke model and the observations, looking for the minimum inside the boundaries $w_{\lambda}=\left\{ 0.001,0.3\right\},\,\, g_{\lambda}=\left\{ -1.0,1.0\right\},\,\, B_{0}=\left\{ 0.2,5\right\}, \text{and}\,\, h=\left\{ 0.0,0.5\right\} $. The uncertainties in the parameters were calculated from the covariance matrix in the minimization algorithm, and they were derived from the square root of the diagonal elements of this matrix. 

\begin{figure*}
    \centering
    \includegraphics[width=0.85\textwidth,angle=0]{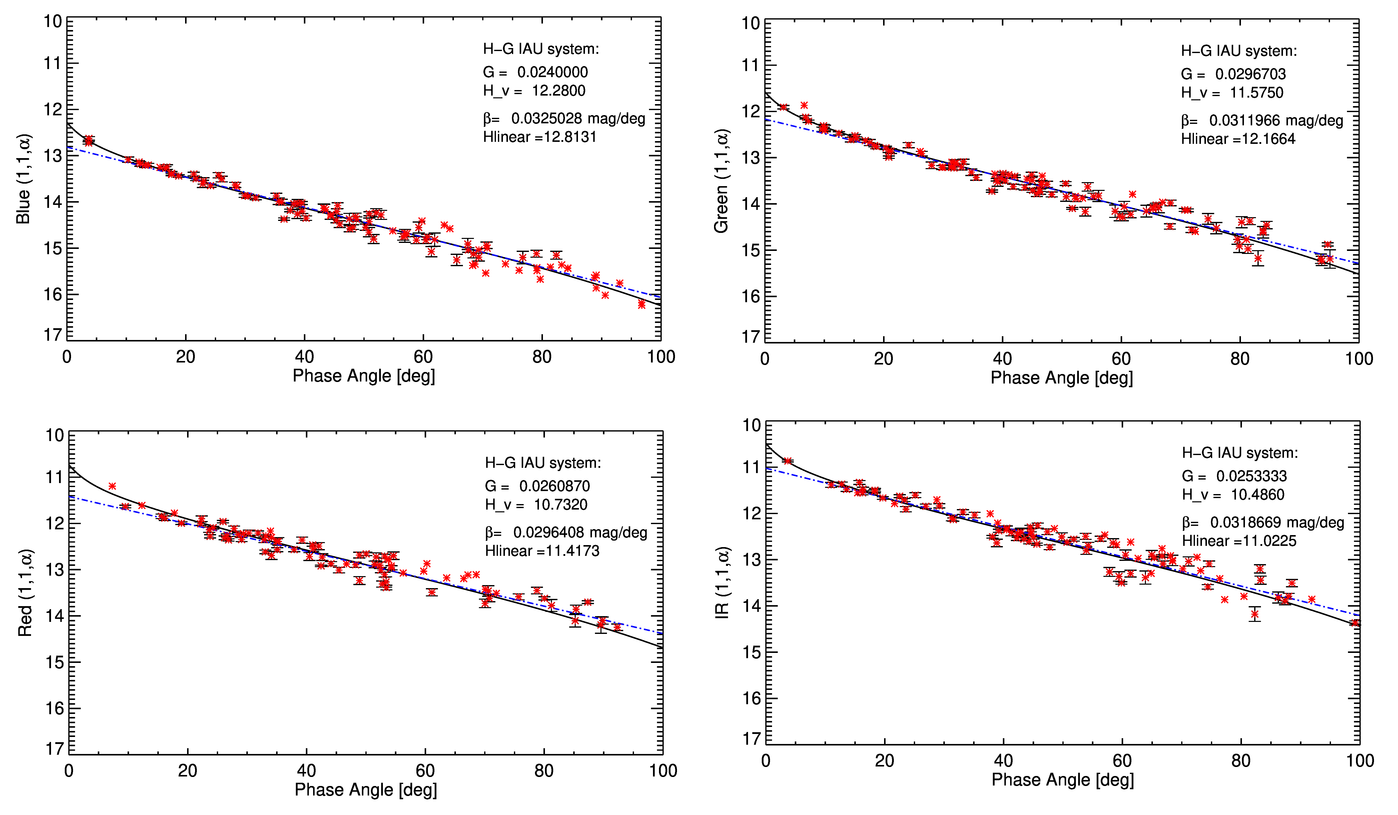}
    \caption{Phobos phase curve in the blue (top-left), green (top-right), red (bottom-left), and IR (bottom-right) filters, respectively. The continuous line represents the HG IAU model that best fits the data, while the blue dashed line represents the linear fit to the data for phase angle $>$ 7 $^{o}$. }
    \label{HGmag}
\end{figure*}
The results are reported in Table~\ref{tab1}.  {Since we fixed the roughness value,  we tested the stability of the results reported in Table~\ref{tab1} for $\bar{\theta}$ ranging from 21$^o$ to 27$^o$, finding that for the lower and higher roughness limits, $w$ and $g$ slightly increase and decrease, respectively, but with values still within the  uncertainties reported in Table~\ref{tab1}}. \\Since the SRC data are not absolutely calibrated, we did not report the SSA in Table~\ref{tab1} to avoid confusion. The estimated fitted value is $w$ = 0.088 for the SRC data, higher than what is obtained for the green and red filters (Table~\ref{tab1}) but lower than the value derived for the IR filter. Considering the very red spectrum of Phobos, this value looks reliable. In any case, these SRC data were used mainly to constrain the opposition surge. We tested the reliability of the derived opposition surge parameters by changing the normalization value to estimate the radiance factor, finding that this factor does not affect the value of B$_0$ and $h$ parameters (changes are within the error bars).   \\
The amplitude and width of the opposition effect are 2.283 $\pm$ 0.029 and 0.05728 $\pm$ 0.0005, respectively. The amplitude of the opposition surge is considerably lower than the one reported by \cite{Simonelli_1998} from the analysis of global-averaged photometry from the Viking mission data acquired with the clear filter in the 1.5-123$^o$ phase angle range (B$_0$= 4$^{+6}_{-1}$, h=0.055$\pm$0.025). These results were determined with very few observations at small phase angles, one observation at $\alpha$ = 1.5$^o$ and a couple at $\alpha$ = 3$^o$, thus originating the large error bars. In the Hapke theory,  B$_0$ was originally expected to vary between zero and one \citep{Hapke_1986}, and later models relaxed this constraint to the zero-to-three range, especially for bodies characterized by irregular surfaces, as commonly found for several asteroids, comets, and planets satellites \citep{Simonelli_1998, Li_2015, Fornasier_2015}.

The SRC data are unique in the coverage of Phobos' small phase angles, and even if not absolutely calibrated, they provide the best estimation of the amplitude and width of Phobos' opposition surge so far. The parameters of the opposition surge of Phobos are very similar  to the ones found for comet 67P from disk-resolved photometry using the Hapke (2012) model \citep{Fornasier_2015}. Cometary nuclei are in fact very red objects, just as Phobos is, and have a comparable albedo, even if their surface is usually even darker (6.5\% at 649 nm for comet 67P). For the disk-averaged photometry of 67P comet, \cite{Fornasier_2015} found an opposition amplitude ranging from 1.91-2.22 in the 480-989 nm range but a width (0.021-0.032) smaller than the one we report for Phobos. 

In Fig.~\ref{fig:SRC2}, we show the observations at the opposition as well as some SRC data acquired at higher phase angles to test the goodness of the model. These data have been corrected for the exposure time and the square of the heliocentric distance and normalized to match the K076 observations at $\alpha$ = 16-18$^o$. Even if the data are scattered, the Hapke fit appears to be good enough, even at high phase angles. \\

To model the HRSC data, we fixed the opposition parameters to the values found with the SRC camera, and we computed the SSA and the asymmetric factor.
We also made an attempt to model the average roughness slope; however, the Mars Express data cover sixteen years of observations and include very different geometry and varying stray light contributions from Mars, resulting in scattering of the data and very large uncertainties in the roughness parameter. Therefore, we decided to fix $\theta$ to 24$^o$, the value determined for the green filter from the disk-resolved photometry. This value is also within the error bars of the roughness determined by \cite{Simonelli_1998} from the Viking data. The results of the modeling are reported in Table~\ref{tab1} and shown in Fig.~\ref{HRSC_allintegrated}. \\
Besides the strong opposition effect, Phobos shows a relatively dark and  backscattering surface. The SSA ranges from 6.4\% in the blue filter to 9.1\% in the IR one, increasing with the wavelength, a behavior expected because of Phobos' red spectrum. The asymmetric {factor} $g$ is negative, indicating backscattering of the incoming light, and slightly decreases at higher wavelengths. The SSA in the green filter is 7.4\%. We compared the values of the reflectance in the green filter with those published by \cite{Pajola_2013} from observations acquired with the OSIRIS cameras on board Rosetta. Their green filter data was centered at the same wavelength as that of Mars Express but narrower, and the authors reported a reflectance value of 0.028 at $\alpha$=$\sim 19^o$, which is very similar to the one we report here that was derived at a similar phase angle (Fig.~\ref{HRSC_allintegrated}). It should be noted that the region observed by Rosetta is in the trailing side and is not observable by Mars Express because it orbits between Mars and Phobos (Mars Express observations cover mostly the leading side of the satellite). Therefore, these similar values indicate that, on a global scale, the reflectance properties of the red unit are very close in the trailing and leading sides. 
\begin{table*}
\centering
\caption{Absolute magnitude of Phobos in the four HRSC filters, following the IAU HG system. The phase integral ($q$), derived from the G parameter, is also reported.}
\label{tab2}
\begin{tabular}{|lcccccccc|}
  \hline
Camera & filter &  $\lambda_c$ (nm)  & $\Delta\lambda$ (nm) & H & G & H$_{lin}$ & $\beta$ [mag deg$^{-1}$] & q\\
  \hline
HRSC &  BL           &  444 & 76  &  12.280 & 0.024 & 12.813 & 0.03250 & 0.3064  \\  
HRSC &  GR           &  538 & 88  &  11.575 & 0.029 & 12.166 & 0.03119 & 0.3102  \\  
HRSC &  RE           &  748 & 48  &  10.732 & 0.026 & 11.417 & 0.02964 & 0.3078 \\   
HRSC &  IR           &  956 & 81  &  10.486 & 0.025 & 11.022 & 0.03073 & 0.3073 \\ 
  \hline
  \hline
\end{tabular}
\end{table*}
When comparing these parameters with those obtained from the disk-averaged photometry of other dark solar system bodies (Table~\ref{tab1}),  Phobos appears quite peculiar, having a stronger opposition effect than the C-type asteroids investigated, excluding Mathilde, for which very few data were available to constrain the opposition surge, and therefore it has large uncertainties in B$_0$ \citep{Clark_1999}. Its steep opposition surge is similar to that of comet 67P and is likely due to a very porous surface that favors a higher shadowing than in dark asteroids. The SSA is higher than that of cometary nuclei or C-type asteroids but comparable to that of the dwarf planet Ceres. The photometric parameters of the distinct Mars terrains reported in the literature are very different from those of Phobos. Mars terrains have much higher SSA values and are less backscattering, and they  usually have a lower roughness and a smaller opposition surge \citep{Johnson_2006b, Johnson_2006a, Jehl_2008}.

\begin{figure*}
    \centering
    \includegraphics[width=0.85\textwidth,angle=0]{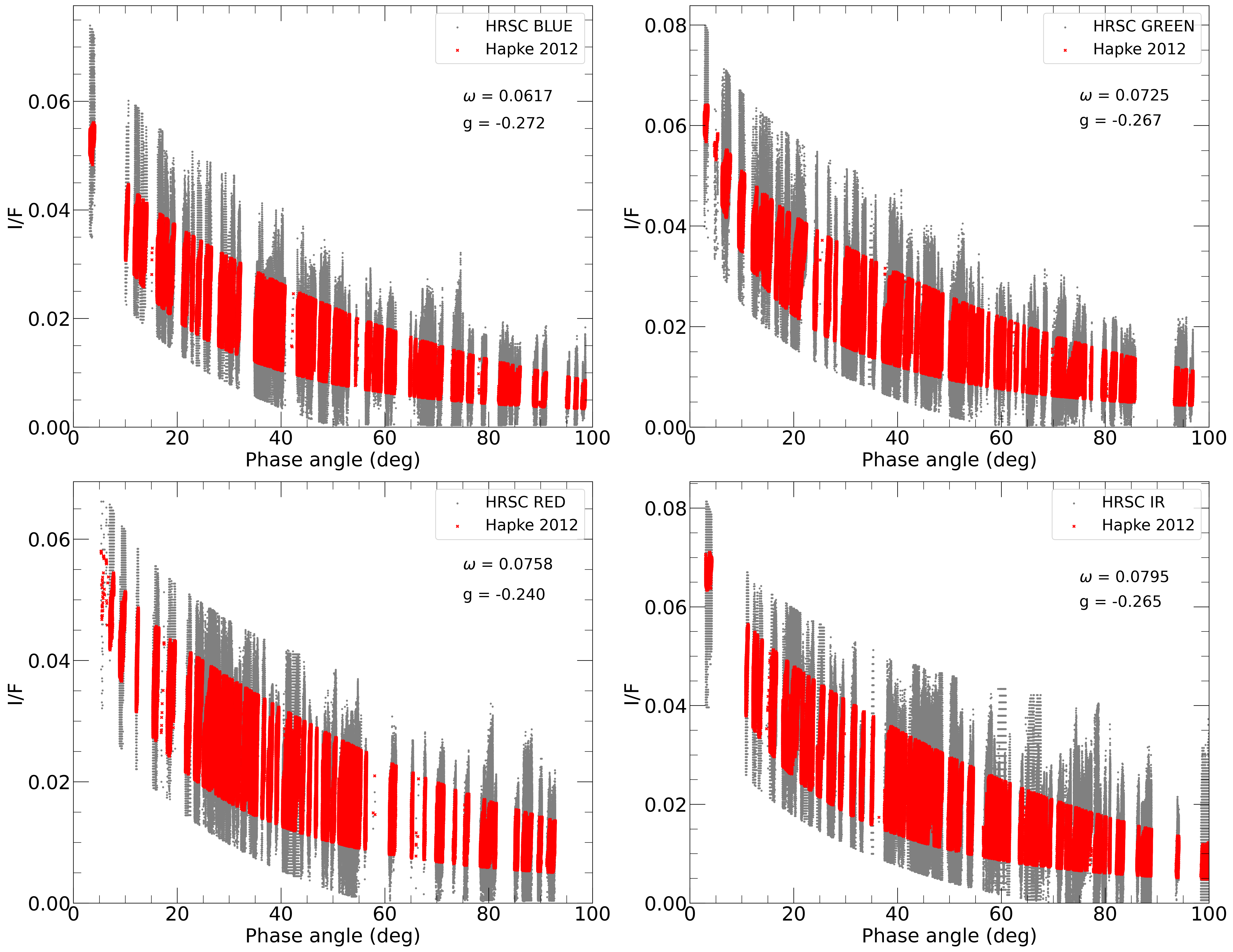}
    \caption{Resolved photometry (gray points) of Phobos with the four calibrated HRSC filters, {where each point represents the reflectance of an element of the Phobos surface at a given incidence, emission, and phase angle, with the modeled data superposed in red~\citep{Hapke_2012a}. These simulations were performed using the 3D shape model \citep{Ernst_2023} and the best-fit solution of the Hapke modeling for each filter (Table~\ref{tab:hapke_res}).}}
    \label{hpk_resolved}
\end{figure*}

\section{Absolute magnitude}

To evaluate the absolute magnitude of Phobos in the four calibrated filters, we first computed the magnitude of the sun in the HRSC filter by convolution of the solar spectrum with the filter transmission. Then, we corrected the flux considering the fact that Phobos' spectrum is redder than that of the Sun using
\begin{equation}
F_{corr} = F_{unc} \times \frac{\int_\lambda F_{\odot}(\lambda)T(\lambda)\mathrm{d}\lambda}{\int_\lambda F_{phobos}(\lambda)T(\lambda)\mathrm{d}\lambda},
\end{equation}
where F$_{unc}$ and F$_{corr}$ are respectively the uncorrected and corrected Phobos reflectance factors at the central wavelength $\lambda_c$ of the considered filter, T($\lambda$) is the filter throughput, and $ F_{phobos}(\lambda)$ and F$_{\odot}(\lambda)$ are the Phobos red unit \citep{Fraeman_2012} and the solar (from the HST catalog) spectra, respectively, both normalized to unity at the central wavelength of the considered filter ($\lambda_c$). \\
We computed the magnitude of Phobos in the given HRSC filter as
\begin{equation}
m_{filt} = -2.5\log \frac{\left(F_{corr} {\int_\lambda F_{phobos} (\lambda)T_V(\lambda) \mathrm{d} \lambda}\right)}{\left({\int_\lambda F_{vega} (\lambda)T_V(\lambda) \mathrm{d} \lambda}\right)},
\end{equation}
where F$_{vega}$ is the flux of Vega from the HST catalog.

The reduced absolute magnitudes in the four filters are shown in Fig.~\ref{HGmag}. We characterized the phase function behavior using the IAU HG system \citep{Bowell_1989}, and the associated results are reported in Table~\ref{tab2}. We also report the linear magnitude (H$_{lin}$), that is, the magnitude computed neglecting the opposition surge; the linear slope coefficient ($\beta$), calculated for $\alpha > 7^o$; and the estimated phase integral $q$ using the equation of \cite{Bowell_1989}:
\begin{equation}
q=0.29+0.684 \times G. 
\end{equation}

The G parameter for Phobos ranges from 0.025 for the IR filter to 0.032 for the blue filter, the averaged value of the phase integral estimated from the G value is 0.308, and the slope parameter $\beta$ is 0.0307$\pm$0.001 mag/$^{o}$. The slope parameter and the phase integral are similar to the values reported in the literature for primordial C-type asteroids. \cite{Veres_2015} found on average a $\beta$ value of $\sim$0.03 mag/deg for C-type asteroids, while \cite{Shevchenko_2019} obtained a phase integral ranging from 0.031 to 0.037 for dark asteroids. The phase integral that we determined is very close to the value reported for Phobos by \cite{Simonelli_1998} (q=0.30$\pm$0.04), while \cite{Shevchenko_2019} found a higher value (q=0.38$\pm$0.03) using the HG1G2 system. The slope parameter is similar to what is already reported in the literature (for instance $\beta$=0.032 and 0.033 mag/deg reported by \cite{Noland_1976} and \cite{Simonelli_1998}, respectively). \\ 
The G value for the primordial asteroids varies depending on the sources. On average, it is 0.09 for D-type asteroids \citep{Veres_2015}, while for C-type asteroids it ranges from 0.07 \citep{Shevchenko_1998} to 0.15 \citep{Veres_2015}, and \cite{Ciarniello_2017} reported a G value of 0.02 for Ceres. Phobos thus has a moderate brightness dependence on the phase angle; this dependence is less steep than the one of comets, which have very similar spectral behavior in the visible range. For instance, comet 67P/Churyumov-Gerasimenko has a negative slope value (-0.13$\pm$0.01) and a steeper phase coefficient ($\beta$= 0.047$\pm$0.002 mag/$^{o}$, \cite{Fornasier_2015}).


\section{Disk-resolved photometry with HRSC}

\begin{table*}[ht]
\centering
\caption{Hapke 2012 parameters derived from the Phobos disk-resolved photometry using the calibrated HRSC data, with $B_{0}$ and $h$ fixed to the values determined from the analysis of the SRC images.}
\label{tab:hapke_res}
\resizebox{\textwidth}{!}{
\begin{tabular}{|ccccccccccc|}
  \hline
  Filter & $\omega$ & g & $B_{0}$ & $h$ & $\bar{\theta}$ (deg) & K & Porosity & Geo. albedo & Bond albedo & Phase integral\\
  \hline
   BL &  0.0617 $\pm$ 0.0011 & -0.272 $\pm$ 0.005 & 2.283 & 0.05728 & 26.42 $\pm$ 1.37 & 1.19 & 0.87 & 0.0611 & 0.0133 & 0.218\\
   GR &  0.0725 $\pm$ 0.0011 & -0.267 $\pm$ 0.005 & 2.283 & 0.05728 & 24.07 $\pm$ 1.26 & 1.19 & 0.87 & 0.0683 & 0.0157 & 0.231\\
   RE &  0.0758 $\pm$ 0.0014 & -0.240 $\pm$ 0.007  & 2.283 & 0.05728 & 21.92 $\pm$ 1.27 & 1.19 & 0.87 & 0.0702 & 0.0165 & 0.245\\
   IR &  0.0795 $\pm$ 0.0018 & -0.265 $\pm$ 0.008 & 2.283 &  0.05728 & 22.36 $\pm$ 1.86 & 1.19 & 0.87 & 0.0770 & 0.0173 & 0.225\\
  \hline
\end{tabular}
}
\end{table*}

The calibrated disk-resolved data obtained from the HRSC camera {were} studied using the latest Hapke IMSA model \citep{Hapke_2012a}. The photometric model was used to characterize the surface through a set of photometric parameters. We created cubes of data containing the original image, phase, emission, incidence, latitude, and longitude images after they had been co-registered following the procedure explained in Section 2.1.
We filtered the data for each cube, excluding the pixels with an incidence and/or emission angle greater than 70$^o$ in order to avoid unfavorable observation geometries near the limb or the terminator. {We also excluded a number of images for which the Phobos shape in the corresponding simulations did not reproduce the original one accurately enough as well as images where the co-registration between the original image and the simulated one was not satisfactory}. 

The Hapke IMSA model \citep{Hapke_2012a} allows fo@500r the estimation of the porosity of the uppermost layers via the porosity factor $K$ defined as:
\begin{equation}
K= - \ln(1-1.209 \phi^{2/3})/(1.209 \phi^{2/3}),
\end{equation}
where $\phi$ is the filling factor and the surface porosity is defined as 1-$\phi$ \citep{Helfenstein_2011}.
In the model, the opposition effect is reproduced considering both the shadow hiding and the coherent backscattering, each having two parameters describing the width and the amplitude of the opposition surge. However, the calibrated data poorly cover the small phase angle range, and it is not possible to correctly evaluate it. For this reason, we constrained the opposition surge using the parameters derived from the SRC camera disk-averaged photometry, therefore considering only the shadow-hiding effect. {Neglecting the coherent backscattering implies underestimating the opposition surge at very small phase angles ($<$ 2$^{o}$). However, it is well known that multiple-scattering is very low for dark surfaces, such as the one of Phobos; therefore, the coherent backscattering should have a negligible contribution in the opposition surge of Phobos}. We modeled all calibrated filter images using a least- $\chi^2$ fit based on the Levenberg-Marquardt algorithm. We looked for solutions within the following boundaries: $w_{\lambda}=\left\{ 0.01,0.3\right\} ,\,\, g_{\lambda}=\left\{ -1.0,1.0\right\}$, and $\bar{\theta}=\left\{ 5^{o},90^{o}\right\} $.  \\
The uncertainties of the Hapke parameters in the disk-resolved case were determined using a bootstrap method { \citep{Efron_1993}.} In fact, the errors determined from the covariance matrix were extremely small because of the large number of points ($>10$ million) fitted  for each filter. In addition, due to the degeneracy in the Hapke model parameters, the fit sometimes converges to a local minimum instead of a global minimum. Therefore, to avoid these problems, we resampled the data, generating smaller sets each including thousands of data in reflectance that we fitted using the Hapke 2012 model. We also resized the boundaries of the parameters closer to the solution found, that is, w$_0$ = 0.06 $\pm$ 0.04, g$_0$ = -0.269 $\pm$ 0.1, and $\theta_0$ = 24 $\pm$ 8 $^{o}$, and we slightly varied the input parameters for each run of modeling. The resampling was performed ten times, resulting in 100 runs of the fit routine with different samples and initial conditions {(examples of the bootstrap data selection and modeling are reported in Fig~\ref{bootstrap}}). The modeling for the different subsets of resampled data gave stable solutions, and finally the parameter values and uncertainties were determined from the mean and the standard deviation of 100 iterations. The best solutions for Phobos disk-resolved photometry are reported in Table~\ref{tab:hapke_res} and shown in Fig.~\ref{hpk_resolved}. \\
Similar to what was found for the disk-averaged solution, the $w_{\lambda}$ increases with the wavelength, though in a less steep way. The asymmetric {factor} indicates backscattering, and its value is almost constant in the different wavelengths, considering the error bars. The roughness parameter varies between 22$^o$ and 26.4$^o$, but with relatively large error bars; therefore, it may be considered almost constant with the wavelength, as expected because it is a geometric parameter in the model and therefore not wavelength dependent. \\
Phobos also has a very high surface porosity (87\%), a value very similar to the one reported in the literature for comet 67P \citep{Fornasier_2015}, indicating that Phobos' top surface is covered by a thick dust mantle with grains that likely have a complex structure, such as fractal aggregates.


\begin{figure*}[t]
    \centering
    \includegraphics[width=0.85\textwidth,angle=0]{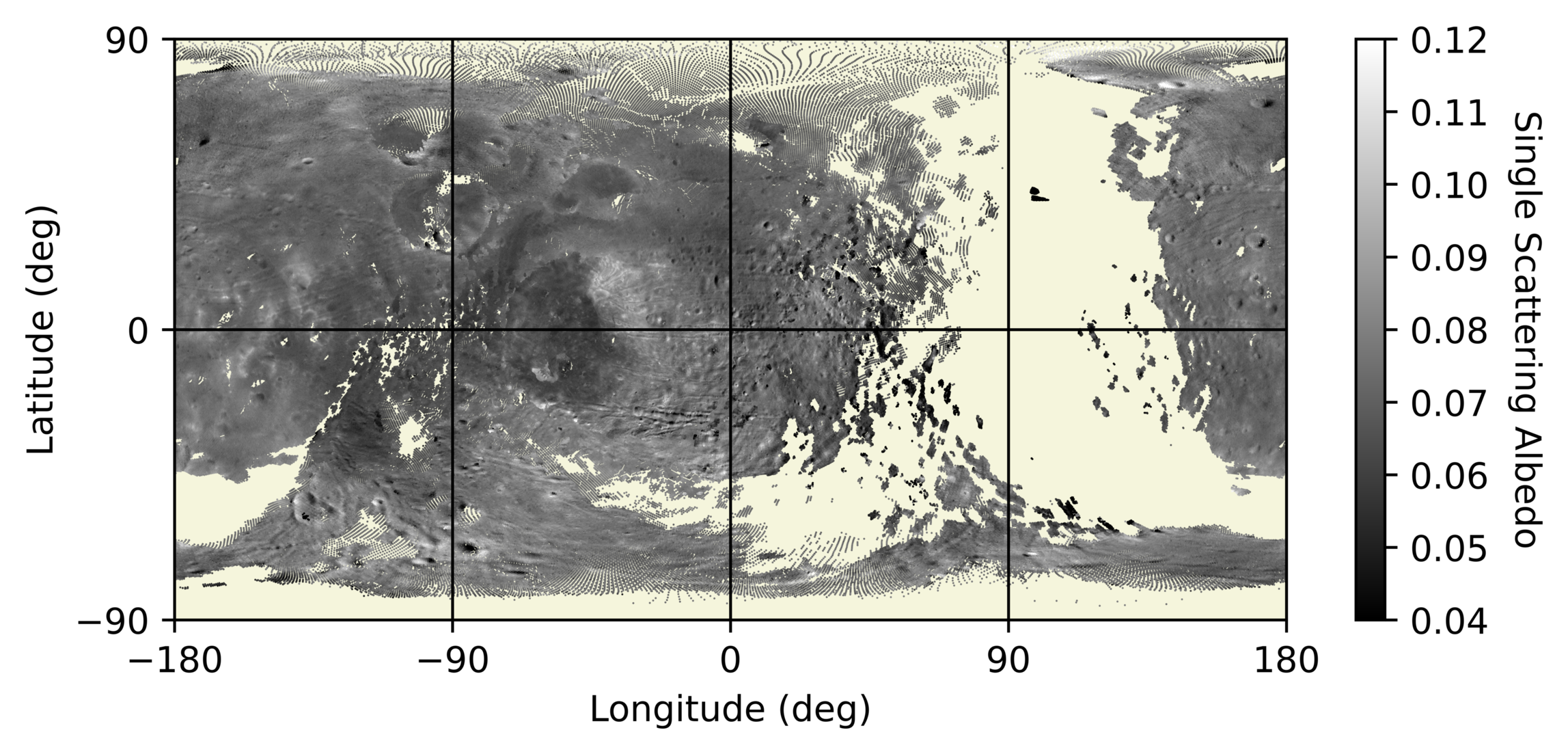}
    \caption{Single-scattering albedo map for the green filter.}
    \label{maps}
\end{figure*}

\begin{figure*}[h]
    \centering
    \includegraphics[width=0.6\textwidth,angle=0]{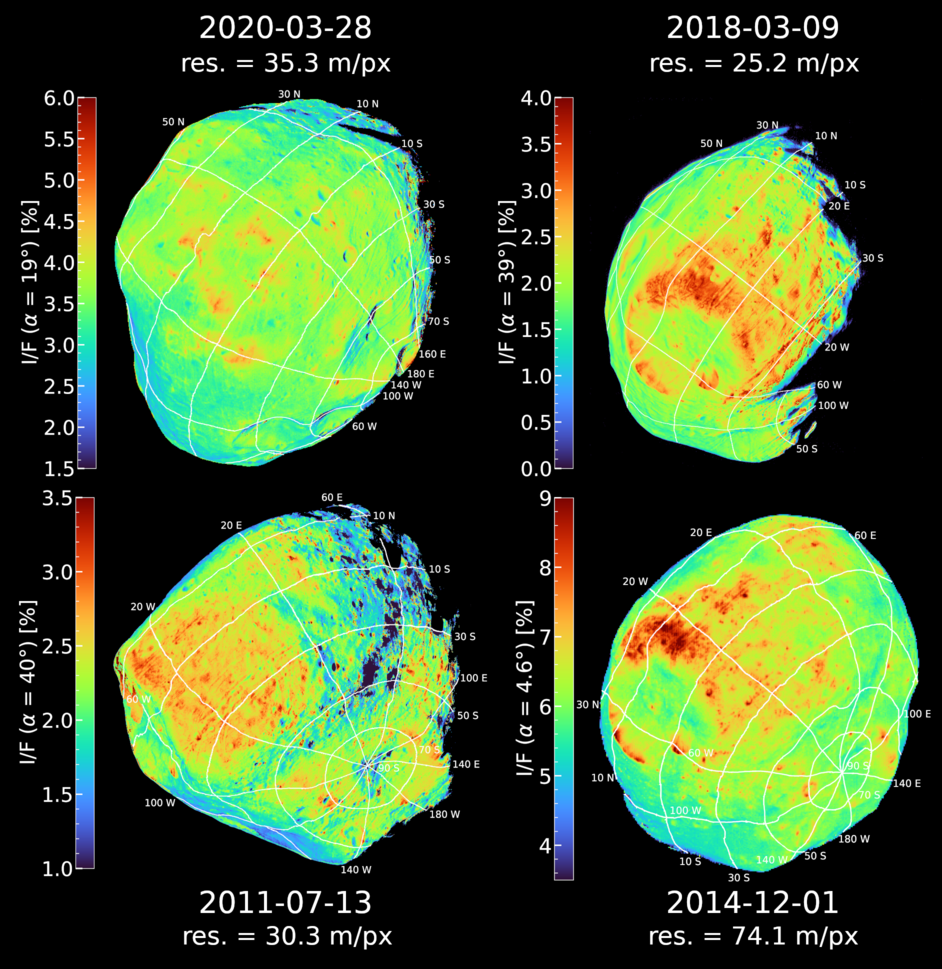}
    \caption{Examples of Phobos' reflectance factor for areas observed at different phase angles. The I/F value has been corrected for the illumination conditions using the Lommel-Seeliger law.}
    \label{IFmaps}
\end{figure*}

\subsection{Single-scattering albedo maps: Phobos reflectance variations}
\label{SSAmaps}

\begin{table*}[t]
\centering
\caption{Hapke 2012 parameters found from the disk-resolved photometry of three regions of interest on Phobos.}
\label{tab:hapke_res_roi}
\small{
\begin{tabular}{|cccccccccc|}
  \hline
  ROIs & Obs. & Latitude & Longitude & $\omega$ & g & $B_{0}$ & $h$ & $\bar{\theta}$ (deg) & Por. (\%) \\
  \hline
   Stickney rim & SRC & 1$\degr$N - 14$\degr$N & 26$\degr$W - 32$\degr$W & 0.092$\pm$0.001  & -0.267 & 2.02$\pm$0.05 & 0.062$\pm$0.001 & 24 &\\
   Limtoc & SRC &8$\degr$S - 15$\degr$S & 57$\degr$W - 61$\degr$W & 0.087$\pm$0.003 & -0.267 & 2.01$\pm$0.07 & 0.056$\pm$0.001 & 24 & \\
   Stickney floor & SRC& 5$\degr$S - 10$\degr$N & 40$\degr$W - 60$\degr$W & 0.059$\pm$0.001 & -0.267 & 2.65$\pm$0.03 & 0.061$\pm$0.001 & 24 &\\
  \hline
     ROIs & Obs. & p$_v$ (\%) & Bond alb. (\%) & $\omega$ & g & $B_{0}$ & $h$ & $\bar{\theta}$ (deg) & Por. (\%) \\ \hline
   Stickney rim & Green & 8.37$\pm$0.05 & 1.87$\pm$0.02 & 0.085$\pm$0.001  & -0.292$\pm$0.004 & 2.02 & 0.062 & 20.6$\pm$1.3 & 86.3 \\
   Limtoc & Green & 7.38$\pm$0.07 & 1.63$\pm$0.02 & 0.075$\pm$0.001 & -0.294$\pm$0.005 & 2.01 & 0.056 & 24.4$\pm$1.6 & 87.5 \\
   Stickney floor & Green & 6.57$\pm$0.05 & 1.18$\pm$0.02 & 0.055$\pm$0.001 & -0.296$\pm$0.005 & 2.65 & 0.061 & 18.2$\pm$1.8 & 86.5\\
 \hline
\end{tabular}
}
\tablefoot{The SRC data were acquired during the orbit numbered K076. For the SRC data photometric model, $\bar{\theta}$ and $g$ were fixed to the value found from HRSC disk-resolved analysis in the green filter (Table~\ref{tab:hapke_res}), while for the HRSC green filter data model, the opposition parameters were fixed to the value determined with SRC. The latitude and longitude coordinates of defined ROIs are also reported for the SRC observations, and the ROIs are the same in the green filter. The term $Por.$ indicates the surface porosity, and $p_v$ indicates the geometric albedo.}
\end{table*}

In the next step, SSA maps were produced in all four HRSC filters. The SSA maps were created by reversing the Hapke (2012) model equation:
\begin{align}
    \omega &= \frac{I/F}{\frac{K}{4} \cdot S(i,e,\alpha, \bar{\theta})} \nonumber\\
    & \times \frac{1}{P_{hg}(\alpha, \omega) \cdot B_{sh}(\alpha, B_{sh,0}, h_{sh}) + B_{cb}(\alpha, B_{cb,0}, h_{cb}) \cdot M \left(\frac{\mu_0}{K}, \frac{\mu}{K}, \omega \right)}.
\end{align}

The set of photometric parameters are from the solutions of the resolved photometric analysis previously determined in each filter. To generate the SSA maps, we carefully selected the images, ensuring good spatial resolution and phase angles that were not too large in order to avoid the important shadowing effect in the analysis. It is worth noting that the trailing side is only partially visible from Mars Express due to the fact that this spacecraft is in a polar orbit positioned between Mars and Phobos. The images used for the maps are listed in Tables B.1, B.2, B.3, and B.4. For overlapping areas observed in more than one image in a given filter, we selected the best resolved image and cropped it accordingly. The blue and green filters have the best image quality in terms of resolution and surface coverage compared to the red and IR filters. The final SSA maps are presented in Figs.~\ref{maps}, and ~\ref{maps_brIR}.\\

The SSA maps correctly display the main geomorphological features of Phobos, including the Stickney and Limtoc craters and the Kepler dorsum (Fig.~\ref{maps}). Considering the map generated with the green filter, the average SSA ($\omega$) is 7.1\%, its value ranges from 5.5\% to 8\% for the red unit, and the blue unit is distinctly brighter, as previously observed in the literature \citep{Fraeman_2014, Simonelli_1998, Pajola_2013}. Notably, the northeast rim of the Stickney crater exhibits a higher reflectance than any other region on Phobos, with $\omega$  higher than 10\%. Some crater rims in the west side of the Stickney crater are also brighter, with $\omega$ $\sim$ 0.010 in the green filter. The floor of the Stickney crater belongs to the darkest area, with $\omega$ $\sim$5.5-6\%. The regions located on the  western part of Stickney are characterized by an SSA value that is between the typical red unit and the darkest region in the Stickney floor (6-6.5\%). \\All of these surface albedo heterogeneities are also evidenced in the images corrected for the illumination conditions using the Lommel-Seeliger law and without applying any phase angle corrections (Fig.~\ref{IFmaps}).

\begin{figure*}
    \centering
    \includegraphics[width=0.86\textwidth,angle=0]{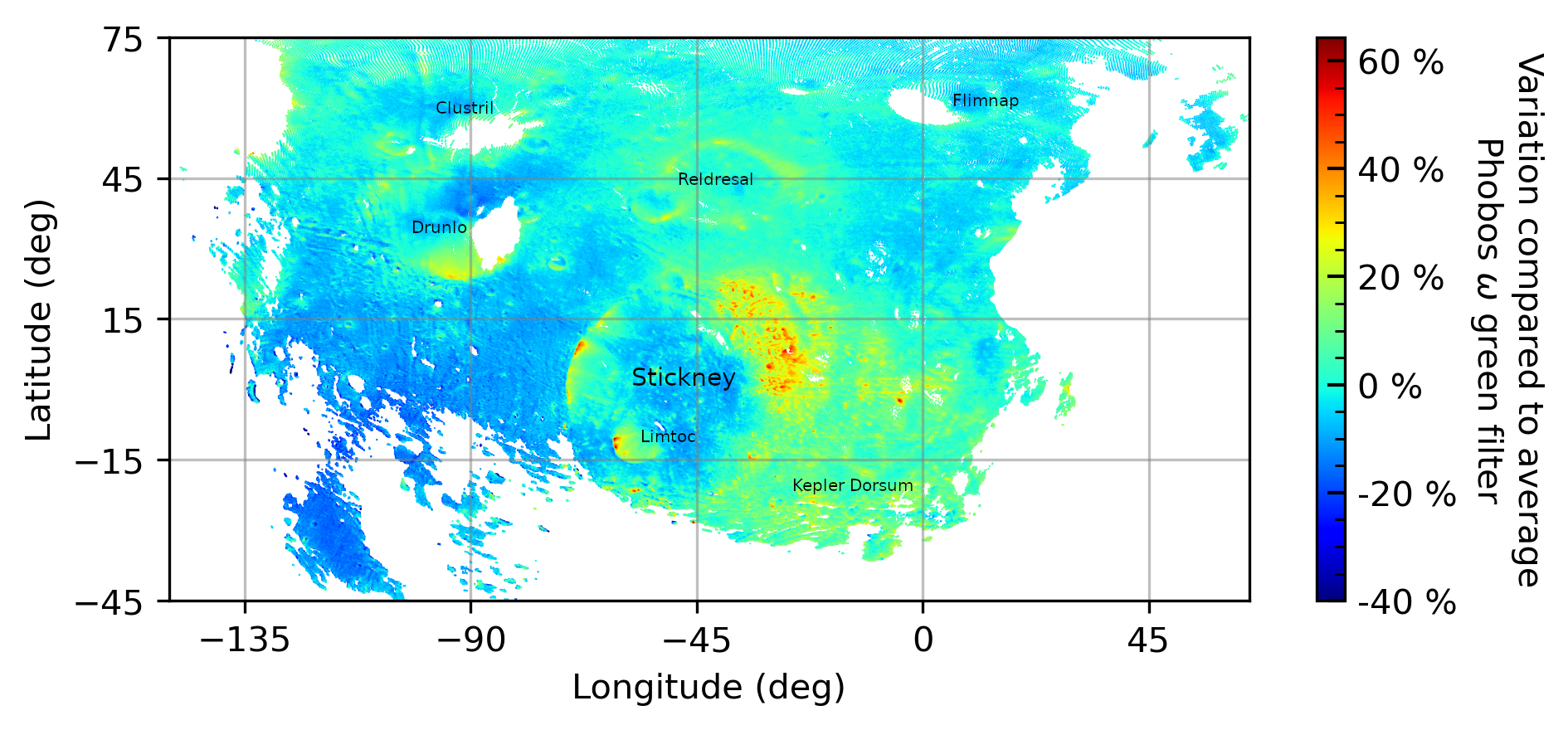}
    \includegraphics[width=0.86\textwidth,angle=0]{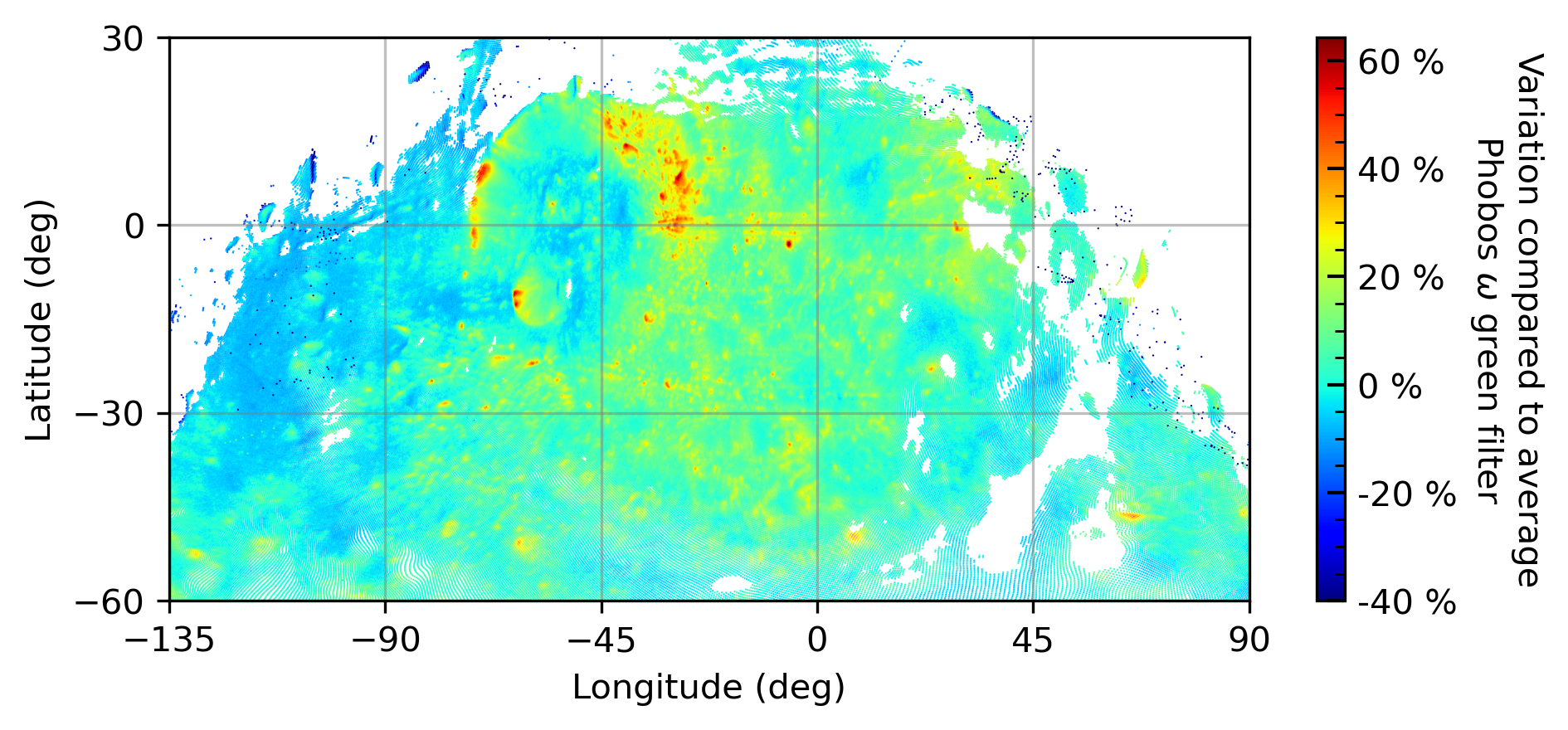}
    \caption{SRC map showing the relative variation of the albedo compared to the average value (7.1\%). Top: Relative albedo variation from the image 2019-11-17T03:27:01 acquired during K076 orbit. Bottom: Relative albedo variation from the image 2020-09-26T02:18:40 acquired during L155 orbit.}
    \label{mapsrc}
\end{figure*}

To better study local reflectance heterogeneities, we also considered the SRC data in SSA relative to the mean value of 7.1\%, which is the average $\omega$ in the green filter. Figure~\ref{mapsrc} maps the relative $\omega$ variations from one image collected during the K076 orbit in November 2019 at a spatial resolution of 22 m~px$^{-1}$ and one from orbit L155 acquired in September 2020 at a resolution of 32 m~px$^{-1}$. Comparing these maps with the main geomorphological features and color units reported in \cite{Basilevsky_2014}, we confirm that the areas dominated by the blue unit have a higher reflectance. The brightest areas of Phobos, with  $\omega$ 50-60\% higher than average, can be found in the northeast Stickney rim, where the blue unit dominates, and observed in relatively small craters producing blue material ejecta as well as on the rims of some grooves. The northwest rim of the Stickney and Limtoc craters also are among the brightest areas, with $\omega$ being locally 40-50\% brighter than average. The south rim of the Drunlo crater, the rims of the Reldresal crater, and the blue unit near the Kepler dorsum are moderately brighter, with $\omega$ 15 to 30\% higher than average, while the regions south of Stickney are 10-20\% brighter. Darker areas can be found near the west side of Stickney, in the equatorial region, where $\omega$ is 10 to 20\% darker than average, and in the hummocky terrain characterizing the floor of Stickney, which is on average 10\% darker and associated with landslides \citep{Basilevsky_2014} but shows locally brighter spots. Grooves usually have a darker floor and brighter rims. More specifically, the grooves located in the blue unit display brighter southern rims. \\

\subsection{Photometric properties of blue and red units}

We also investigated the variations in the opposition effect among three different regions of interest (ROIs) on Phobos, two representing the blue unit (northeast rim of Stickney and the brightest rim of the Limtoc crater) and one in the red unit located in the floor of Stickney crater (Fig.~\ref{locationROI}). We first analyzed the SRC images since they provide the best coverage of the Phobos opposition effect, especially around the Stickney crater, where the images are particularly well-resolved (spatial resolution of 23 m~px$^{-1}$) at low phase. The K076 orbit is especially noteworthy due to its phase angle coverage, ranging from 1 to 17$^{o}$. Since these data are not absolutely calibrated, we applied the same calibration factor used for SRC disk-integrated analysis. Because we only have data at small phase angles, we set $\bar{\theta}$ to 24$^{o}$ and g to -0.267 in the SRC ROI-resolved photometric analysis. The asymmetry parameter value was chosen based on the results of the disk-resolved Hapke fitting for the green filter. The results of the SRC disk-resolved analysis (Table~\ref{tab:hapke_res_roi}) confirmed the main findings previously discussed: the Stickney crater's northeast rim covered by blue unit material is significantly brighter ($\omega$ = 0.092) compared to the crater's floor ($\omega$ = 0.059), resulting in an $\omega$ difference of $\sim$56\%. Apart from the reflectance differences, the phase functions of the different ROIs of Phobos (and consequently of the different Phobos units) exhibit diverse behaviors. In particular, the B$_0$ parameter appears slightly lower for the blue unit (Stickney rim, B$_0$ = 2.02) than for the red unit (Stickney floor, B$_0$ = 2.65). Therefore, the opposition effect is more pronounced for the red unit than for the blue one. \cite{Simonelli_1998} previously also found global changes in the phase function among the two units.  \\
These authors also showed that the reflectance of the dark material usually located in the floor of craters tends to darken faster with increasing phase angle than the bright material. The dark material found in the Stickney crater floor and in other smaller craters \citep{Goguen_1978}, located mostly in the trailing side and not observable by Mars Express, have been reported to have lower $\omega$, to be more backscattered, and to be characterized by a higher roughness than Phobos' average terrain \citep{Goguen_1978, Simonelli_1998}. We investigated the asymmetric {factor} and roughness of the three ROIs using the green filter observations. We fixed the opposition parameters to the values determined by SRC observations for each individual ROI, and we ran the Hapke 2012 model, this time leaving $\omega$, $g$, and $\bar{\theta}$ as free parameters. Indeed, we found that the SSA at 538 nm is lower in the red unit material composing the floor of Stickney. However, the ROIs located in the blue and red units have the same asymmetric factor value, indicating very similar backscattering properties (Table~\ref{tab:hapke_res_roi}). Additionally, we did not observe differences in the surface porosity for the red and blue unit ROIs investigated in this work. The porosity has in fact a high value indistinguishable from the Phobos' average one reported in Table~\ref{tab:hapke_res}. Concerning the roughness, our results do not support the conclusions of \cite{Goguen_1978} and \cite{Simonelli_1998} that the floor of craters is significantly rougher than the average terrain, at least for the floor of Stickney. In fact, our analysis shows that it has a roughness parameter of 18.2$^o$, which is slightly lower but comparable within the uncertainties to the value found for the Stickney rim and is significantly lower than the roughness found in the bright landslide of the Limtoc crater (Table~\ref{tab:hapke_res_roi}). Stickney and Limtoc are among the oldest regions on Phobos, with ages ranging from 2.6 Gyr, if Phobos is a captured asteroid, to 4.3 Gyr, if Phobos was formed in the present orbit \citep{Schmedemann_2014}. The ROIs representing the blue unit we investigated include the Stickney ejecta and the bright landslide in Limtoc, therefore potentially relatively younger areas. Our analysis indicates that the blue and red units not only have different spectral behaviors and brightness, but they also have different opposition parameters, with the darker material in the floor of Stickney having a steeper opposition effect than the blue unit ROIs but sharing a similar backscattering behavior. The Stickney floor, resulting from accumulation of downslope material movement, appears to be dominated by darker particles and is smoother than the average Phobos terrain. We note that this area also includes brighter spots of blue material not fully resolved in the images used in this analysis.

\section{Discussion}

Phobos shows important heterogeneity represented not only by the two distinct spectral units (blue and red) but also in the albedo. The brightest regions, generally found on Stickney's northeast rim, can have a reflectance 50-65\% higher than the average terrain, while darker material is mostly found inside craters and in the regions located at midlatitude and on the dark red unit located at the west side of Stickney.\\
The selected ROIs within the blue and red spectral units display distinct photometric characteristics. Specifically, the red unit exhibits a higher opposition surge and a lower reflectance compared to the brighter blue unit. In what follows, we compare the Stickney crater properties with those of craters on other small bodies observed by space missions. The Phobos surface is under the gravity dominated regime (diameter $\gtrsim10$ km), where cratering generally produces roundish craters with well-defined ejecta blankets and rims~\citep{Holsapple_1993}. Other objects under the same regime and recently explored by space missions are (2867) Steins (ESA/Rosetta), (21) Lutetia (ESA/Rosetta), (4)
Vesta (NASA/DAWN), and (1) Ceres (NASA/DAWN) and, in the Kuiper belt, the classical Kuiper belt object (486958) Arrokoth, observed by NASA's New Horizons spacecraft. Even though these bodies are very different in composition, formation processes, and evolution, we try to qualitatively compare the photometric properties of the craters observed on them. \\
From the aforementioned sample, only Ceres and Arrokoth have primitive carbonaceous organic-rich analog compositions. Ceres is a 965-km
dwarf planet whose surface is scattered with very defined craters. Occator is the most intriguing crater on the Ceres surface. The
bright central faculae connected to brines and fresher material make this crater particularly unique. On the other hand, Narwish and Kerwan quadrangles are considered to
be much older ($\sim2$ Ga, \cite{Carrozzo_2019} and $\sim1$ Ga, \cite{Williams_2018}). From the maps of the photometric parameters produced by \cite{Li_2019}, the Narwish region
is characterized by a lower $\mathit{{w}}$, stronger $g$, and higher $\bar{\theta}$ than Ceres. Kerwan also displays a high $\bar{\theta}$ but
lower $g$ and higher $\mathit{{w}}$. Narwish's photometric parameters are similar to the characteristics of the Stickney floor, where $\mathit{{w}}$
is proportionally lower and $B_{0}$ is higher. While $B_{0}$ and $\bar{\theta}$ do not probe the same size scale, a higher $B_{0}$ value in dark bodies is generally related to rougher and irregular particles and to the increase of mutual inter-particle shadows. \\
For (486958) Arrokoth, which has approximately a size similar to that of Phobos but a complex bilobated shape, observations are available in only three different phase angles (12$^{o}$, 16$^{o}$, and 33$^{o}$), thus limiting the comparison to only albedo variation.
The largest noticeable crater is found in the smallest lobe and is called Sky. Sky shows two apparent landslides at its crater walls, landslides which are 60\% brighter than the overall Arrokoth \citep{Hofgartner_2021}. The crater floor is partially hidden, but it is apparently neutral or slightly darker. Phobos also
shows bright landslides in the eastern walls of Stickney and Limtoc,
reaching 40--60\% higher reflectance than the average terrain. Landslides in crater walls are apparently
ubiquitous, and they are also present in (21) Lutetia, (4) Vesta, and Ceres, but they are darker in the Lutetia Baetica region \citep{Hasselmann_2016} and varied in Ceres
as well as Vesta. Dark landslides have been proposed to be an outcome of
particle sorting \citep{Hasselmann_2016}: small particles are levitated
during the event, thus becoming lost or transported away. Bright landslides,
on the other hand, are a possible outcome of fresher materials or diverse
subsurface composition becoming exposed, as observed on Phobos.

\begin{figure}
    \centering
\includegraphics[width=0.46\textwidth,angle=0]{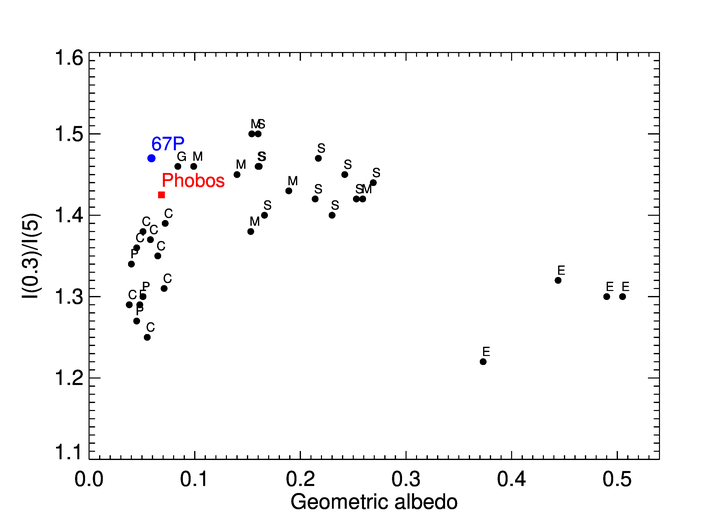}
\includegraphics[width=0.46\textwidth,angle=0]{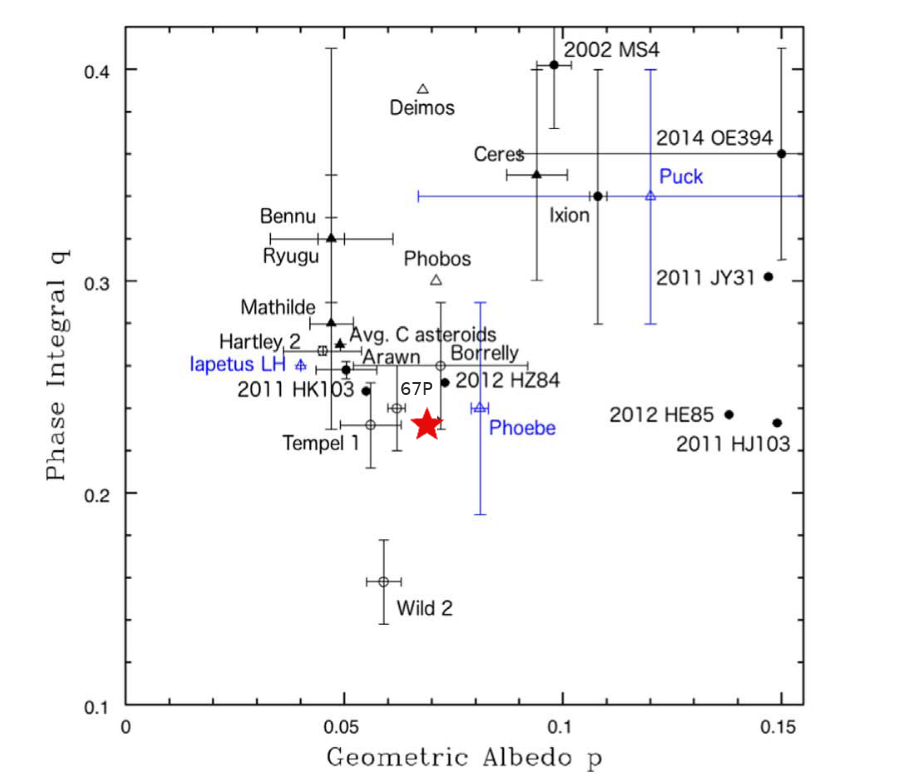}
 \caption{Comparison of Phobos photometric properties with those of other Solar System bodies. Top: Phobos reflectance ratio (at $\alpha$=0.3$^o$ over $\alpha$=5$^o$) versus geometric albedo compared to the values observed for different classes of asteroids (following the \cite{Tholen_1989} classification scheme), derived from \cite{Belskaya_2000}, and comet 67P \citep{Fornasier_2015}. Bottom: Figure adapted from \cite{Verbiscer_2022} showing the phase integral versus  geometric albedo for dark Solar System bodies. The Phobos value derived in this paper from the green filter and using the Hapke modeling of resolved photometry is represented by a red star (uncertainties are within the star symbol). Literature data of Phobos (q=0.30) from \cite{Simonelli_1998} are also reported. We refer to \cite{Verbiscer_2022} (their Fig. 11) for the references of the individual phase integral and geometric albedo values of the objects.}
    \label{compara}
\end{figure}

\begin{figure}
    \centering
\includegraphics[width=0.46\textwidth,angle=0]{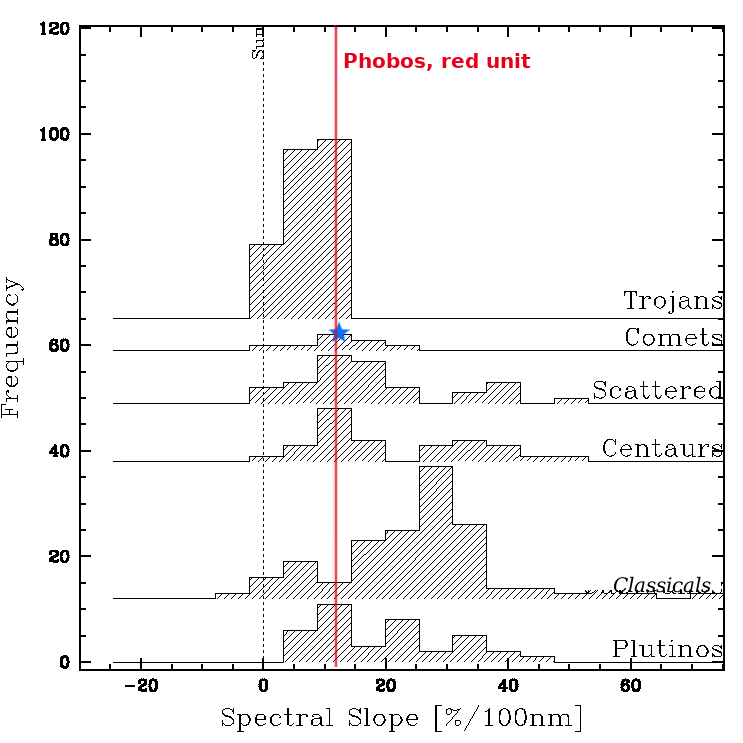}
 \caption{Spectral gradient, estimated in the 550-800 nm range, of Phobos' red unit (11.9 \%/(100 nm; red vertical line) derived from spectra presented in \cite{Fraeman_2012} compared to the values reported by \cite{Fornasier_2007} for cometary nuclei, Jupiter Trojans, and the different dynamical classes of the transneptunian population. Frequency refers to the number of objects in the different histograms. The figure was adapted from \cite{Fornasier_2007}. The blue star indicates the spectral slope of comet 67P determined at phase 1.3$^{o}$ (12.6\%/(100 nm)) and a heliocentric distance of 3.6 au by \cite{Fornasier_2015}.}
    \label{comparatnos}
\end{figure}

We also attempted to compare Phobos' photometric properties with those of meteorites and minerals. The bidirectional distribution functions (BRDF) measured in the laboratory on different materials, grain sizes, and geometric conditions demonstrate that it is difficult to obtain a phase function similar to that observed on Phobos. In fact, the scattering function is usually more isotropic in the laboratory \citep{Souchon_2011, Johnson_2013, Potin_2022, Wargnier_2023}, resulting in a significant increase in reflectance after a 80-100$^o$ phase angle due to forward scattering. Moreover, laboratory phase function measurements indicate that the opposition surge is often less significant than expected or observed for Solar System bodies. For instance, \cite{Beck_2012} found a B$_0$ parameter of approximately 0.4 for the Tagish Lake meteorite, whereas observations of primitive asteroids, from which the meteorite could have originated, typically exhibit values greater than 1.5. None of the different meteorites measured by \cite{Beck_2012} (including carbonaceous chondrites, Tagish Lake meteorite, howardites, eucrites, diogenites, and a lunar sample) show an opposition surge as strong as the one observed for Phobos. If we computed the absolute and relative opposition effect intensity (OEI) from the SRC observations of Phobos using the same definition as \cite{Beck_2012}\footnote{The OEI is the difference of the reflectance at 0 and 30$^o$ of phase angle, while the relative OEI is the ratio of the reflectance at 0$^o$ over that at 30$^o$}, we obtain for Phobos a value of $\sim$ 0.06 for the absolute OEI and $\sim$ 3.7 for the relative OEI, values that are much higher than what was reported for the meteorites measured by \cite{Beck_2012} (see their Fig. 6). It should be noted that the reflectance at zero phase angle is often extrapolated in laboratory data, where measurements usually cannot be performed at phase angles smaller than $\sim$ 5$^o$ and different parameters such as grain size and shape, surface rugosity, and porosity affect the shape of the phase function \citep{Okada_2006, Kamei_2002}. For instance, studies conducted on different grain sizes of the same sample have shown that the forward scattering tends to be larger for the finer grains \citep{Kamei_2002, Johnson_2013}, and that larger particles tend to diffuse more backward \citep{Souchon_2011}. Moreover, for smaller grains, the effect of surface charging on particle scattering can also induce discrepancies between laboratory and remote sensing data \citep{Klacka_2007}.

Overall, our photometric analysis has shown that Phobos photometric properties {show a close resemblance} to those of comet 67P: both have a red spectrum, a high surface porosity, and similar opposition effect values, even though it should be noted that the material on top of comet 67P has a lower SSA (3.4-4.2\%) compared to Phobos (7.4\%), and it is more backscattering (g=-0.42 for 67P, and -0.27 for Phobos). As discussed in the previous sections, the SSA of Phobos is higher than other primordial bodies investigated by space missions (Bennu, Ryugu, Mathilde) and comparable to that of the dwarf planet Ceres.\\
Figure~\ref{compara} shows Phobos' reflectance ratio at a small phase angle versus the geometric albedo compared to the properties of different asteroidal classes \citep{Belskaya_2000} and comet 67P \citep{Ciarniello_2015,Fornasier_2015}. Phobos has a strong opposition surge that is usually higher than most carbonaceous asteroids investigated and is comparable in intensity to that of metallic or silicate-rich asteroids. However, its albedo is much lower than the one of silicate or metallic asteroids and similar to that of primordial asteroids. Concerning D-type asteroids, which are characterized by a red spectrum (as also observed for Phobos), there are very few data in the literature. Both \cite{Shevchenko_2012} and \cite{Mottola_2023} reported observations at a small phase angle for Jupiter Trojans (including four D-types), finding that these bodies have a linear dependence in the phase function with almost no opposition surge and usually a very low albedo. 

Actually, comet 67P is the body closest to Phobos in terms of photometric properties when considering both the albedo and opposition surge properties. This similarity is also confirmed when comparing the geometric albedo and the phase integral of different dark Solar System bodies (Fig.~\ref{compara}). In this paper, we have derived a phase integral of 0.23 in the green filter from Hapke-resolved photometry and of 0.31 from the HG fitting of magnitude in the same filter from integrated photometry. In the literature, there are several phase integral estimations of Phobos with quite different values: \cite{Pang_1983} reported q=0.33 and \cite{Klaasen_1979} q=0.27,  while \cite{Shevchenko_2019} reported q=0.38$\pm$0.03 (from HG1G2 model) and \cite{Simonelli_1998} q = 0.3. Deimos has a phase integral of 0.38 \citep{Thomas_1996}. \\
Among the two values reported in our analysis, we think that the one derived from the Hapke-resolved photometry should be more reliable because it is based on data acquired at a very small phase angle, therefore permitting accurate modeling of the Phobos strong opposition effect, while the HG model was based on the reduced magnitudes derived from absolutely calibrated filters, which have a limited coverage of the opposition surge. 
When comparing the phase integral and geometric albedo of Phobos with those of other dark Solar System bodies (Fig.~\ref{compara}), it appears once more that Phobos has photometric properties similar to those of Jupiter family comets, which originate from the Kuiper belt. Phobos' phase integral is also very close to that of Phoebe, which is supposed to be a Kuiper belt object captured by Saturn. Additionally, we also found that the opposition effect of Phobos is very similar in shape and parameters to the one of comet 67P, and it has a similar high surface porosity. 
In Fig.~\ref{comparatnos} we also compare Phobos' red unit spectral slope, derived from \cite{Fraeman_2012} spectra, to that of other primordial bodies, derived from the analysis of \cite{Fornasier_2007}, such as Jupiter Trojans, which are dominated by D-type asteroids; cometary nuclei; and the different dynamical classes of the transneptunian population. The blue star refers to the spectral slope of comet 67P determined at $\sim$ 3.6 au pre-perihelion and at a phase angle of 1.3$^o$ \citep{Fornasier_2015}. As already well known, Phobos has a red spectral behavior close to that of D-type asteroids, and it also shares spectral similarities with cometary nuclei, notably comet 67P; with Jupiter Trojans; and with the moderately red bodies of the transneptunian population.  \\  
Based on the aforementioned similarities in the photometric and spectral properties with primordial Solar System bodies, particularly with cometary nuclei, we suggest that additional dynamical investigations deserve to be performed in order to understand the origin of Mars satellites, mimicking the capture not only of asteroids but of other small bodies. These simulations may consider a bilobated or binary comet or external Solar System body capture to test the hypothesis that Phobos and Deimos may currently be exhausted cometary nuclei captured by Mars. A binary or bilobated comet might eventually be fragmented into two bodies by Mars' tidal and gravitational forces during the capture process and may have subsequently generated the Martian moons.     
  

\section{Conclusions}

We have analyzed photometric observations performed by the Mars Express mission on Phobos with the HRSC absolutely calibrated filters in the blue, green, red, and IR range and with relatively calibrated panchromatic data acquired with the SRC camera. In this work, we have presented the results of the photometric modeling for the different HRSC filters, the SSA maps, and a comparison of the 
photometric properties of Phobos with those of other satellites and dark minor bodies. The data acquired with the SRC camera are unique and represent the best coverage achieved so far of the Phobos opposition effect. This camera in fact acquired {299} images covering the 0.3-17$^{o}$ phase range. The main findings of our work are the following:
\begin{itemize}
\item Phobos has a strong opposition effect characterized by an amplitude B$_0$ =2.283$\pm$0.029 and a width 0.05728$\pm$0.0005.
\item Phobos is a relatively dark object with a mean albedo in the green filter of 6.83\%. The SSA increases with wavelength from 6.2\% in the blue filter to 8\% in the IR filter, as expected because of Phobos' very red spectrum.
\item Its surface is backscattering (-0.027$<g<$=-0.024) and has a high porosity value (87\%), indicating the presence of a thick dust mantle, possibly composed of grains with a  complex shape or fractal aggregates.
\item The SSA maps reproduced the main geomorphological features of the Mars satellite and provide clear evidence of the albedo dichotomy between the blue unit, which is up to 50-65\% brighter than average in the northeast rim of Stickney, and the darker red unit. The darkest regions are found in the floor of Stickney and in the regions located in the western side of this crater. 
\item Local photometric analysis of selected areas located in the blue and red units showed that the red unit terrain has a lower SSA and a stronger opposition surge compared to the blue unit terrain, but they have similar surface porosity values and backscattering properties.
\item The phase function of Phobos shares analogies with dark asteroids, but its SSA is considerably higher than the values reported for carbonaceous-rich asteroids, such as Bennu, Ryugu, and Mathilde, and similar to the value found for the dwarf planet Ceres.
\item The opposition surge parameters, the porosity, and the phase integral of Phobos are very close to the value reported in the literature for the comet 67P, which is characterized by a similarly red spectral slope but has a much lower SSA value.
\item The phase integral of Phobos is similar to that of dark asteroids of the Jupiter family comets and Phoebe. While waiting for detailed in situ analysis of D-type Jupiter Trojans by the Lucy mission, so far the closest analogs to Phobos photometric properties are cometary nuclei. However, the material comprising Phobos has a higher SSA than that on the top surface of comets or primordial asteroids.    
\item Based on the photometric properties that are similar to the comet 67P, we suggest investigating the dynamical capture of a binary or bilobated comet as the potential origin of the Martian moons. 

\end{itemize}

All of these results are of high interest and support to the JAXA MMX mission. The mission's main goals are the return of Phobos samples collected in both the red and blue units, a detailed investigation of Mars' satellites, and determining the origin of the Martian moons.

\begin{acknowledgements}

We acknowledge the ESA Planetary Science Archive for space mission data procurement and the principal investigator of the HRSC instrument, G. Neukum (Freie Universitaet, Berlin, Germany). We thank the anonymous reviewer for valuable comments and suggestions, which helped us to improve the quality of the manuscript.

\end{acknowledgements}

%
%


\bibliographystyle{aa}
\bibliography{references}


%


%

%

\newpage

\begin{appendix}

\section{Supplementary material: Figures}

\begin{figure*}
    \centering
    \includegraphics[width=0.7\textwidth,angle=0]{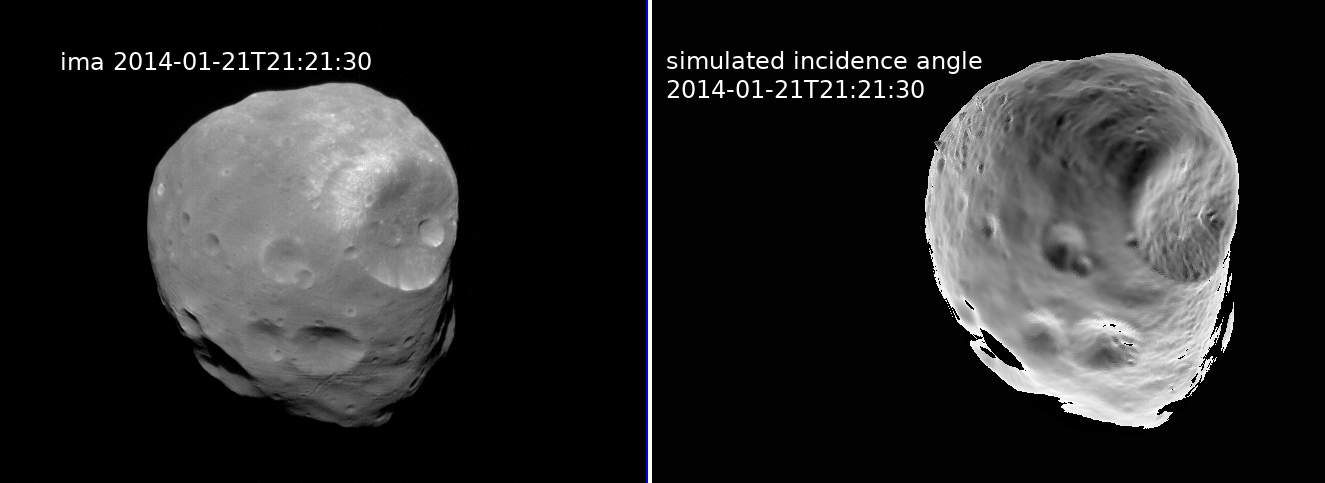}
    \caption{The image on the left/right is an example of original/simulated image (incidence angle). The simulated image is not satisfactory. In fact, Phobos' shape is larger in the simulated image than in the original one, and it was thus discarded from our analysis.}
    \label{shape_nogood}
\end{figure*}

\begin{figure*}
    \centering
    \includegraphics[width=0.46\textwidth,angle=0]{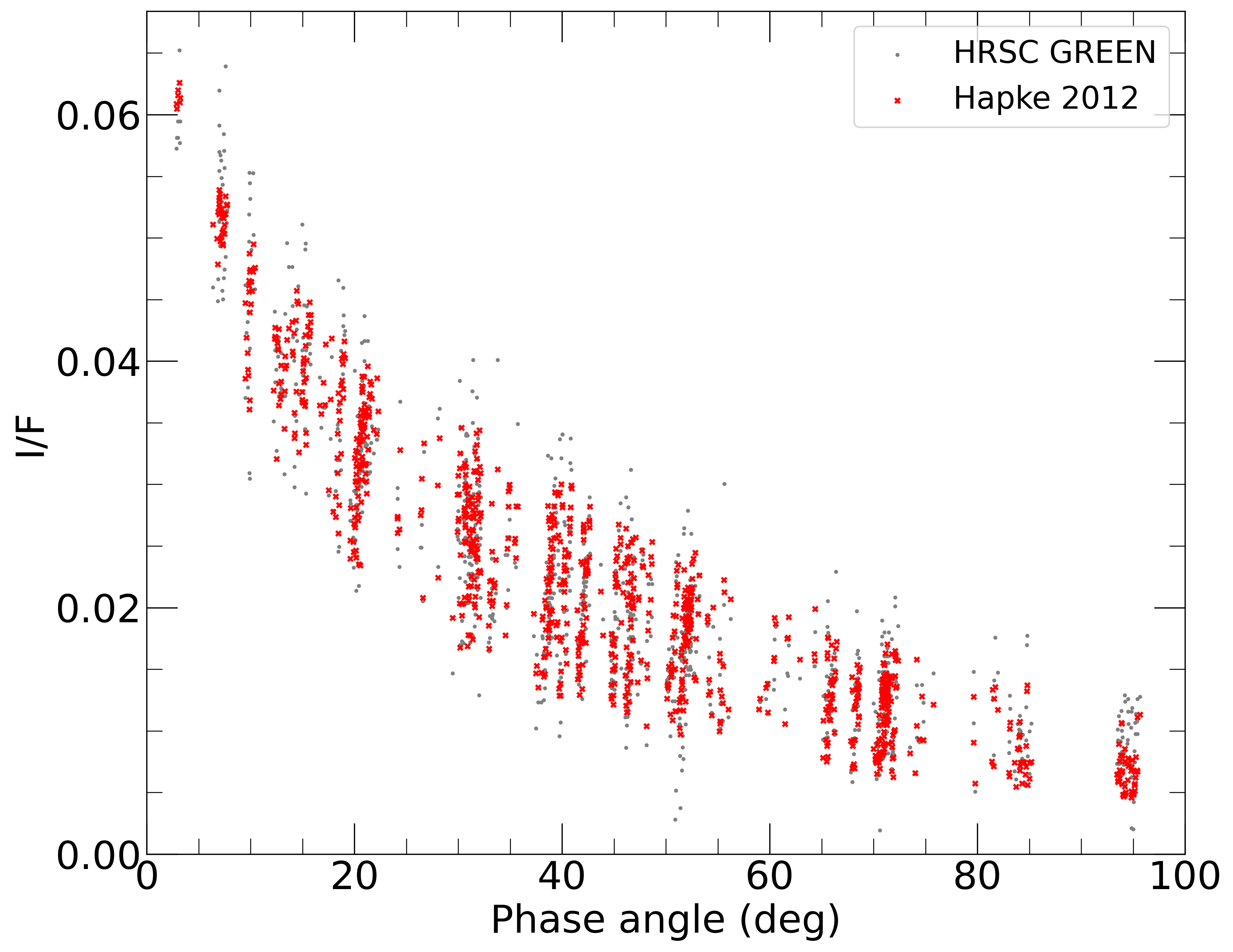}
    \includegraphics[width=0.46\textwidth,angle=0]{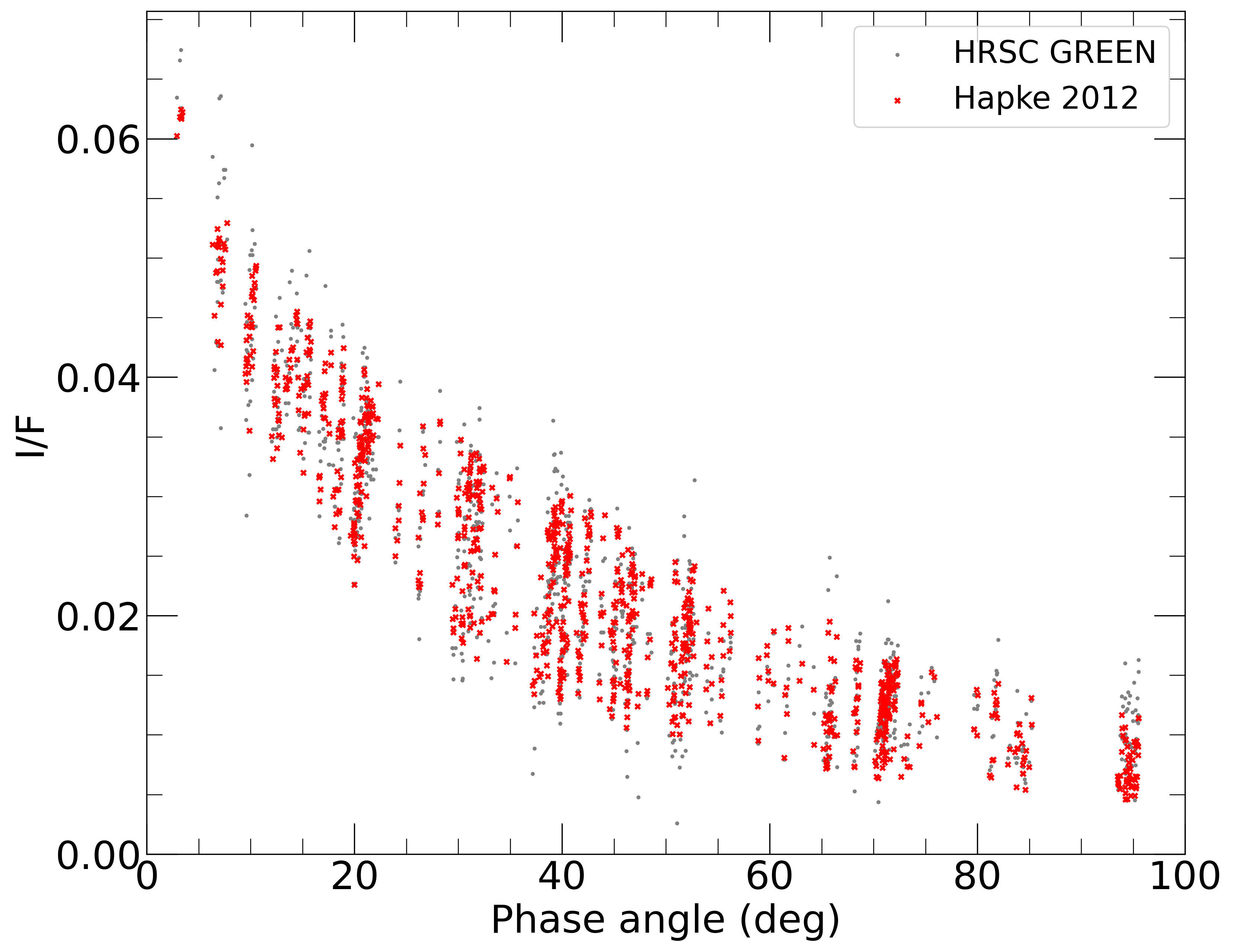}
    \caption{Example of bootstrap method for the green filter resolved photometry. The whole sample, which included several million points, was resampled to generate smaller datasets of approximately one thousand points each. Here, two subsets are shown (gray points) together with the best-fit data models (red points).}
    \label{bootstrap}
\end{figure*}

\begin{figure*}
    \centering
    \includegraphics[width=0.8\textwidth,angle=0]{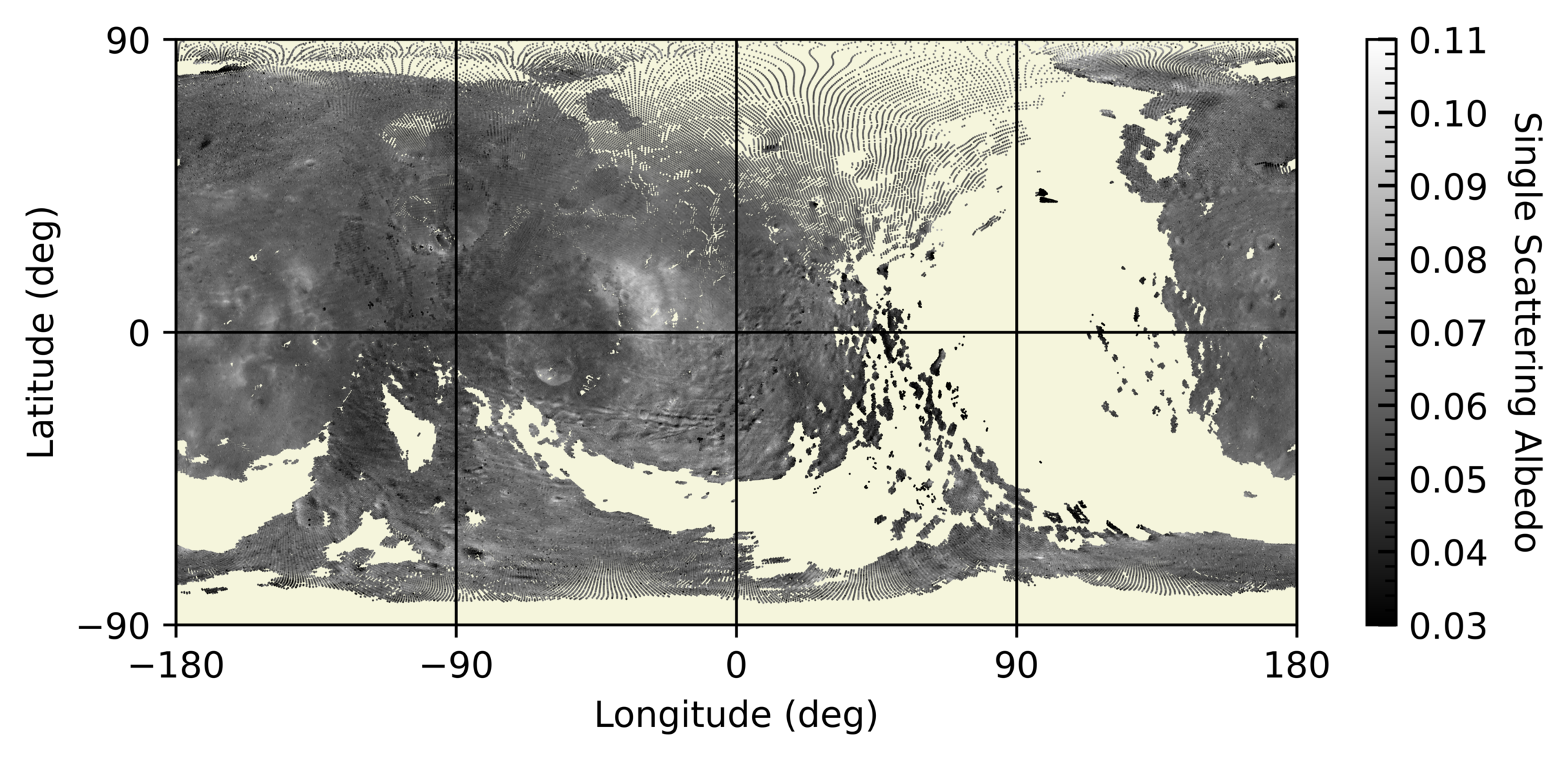}
    \includegraphics[width=0.8\textwidth,angle=0]{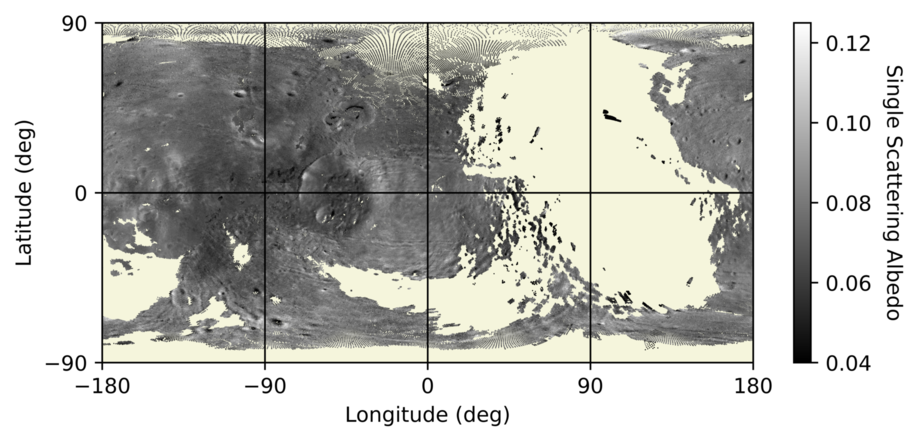}
    \includegraphics[width=0.8\textwidth,angle=0]{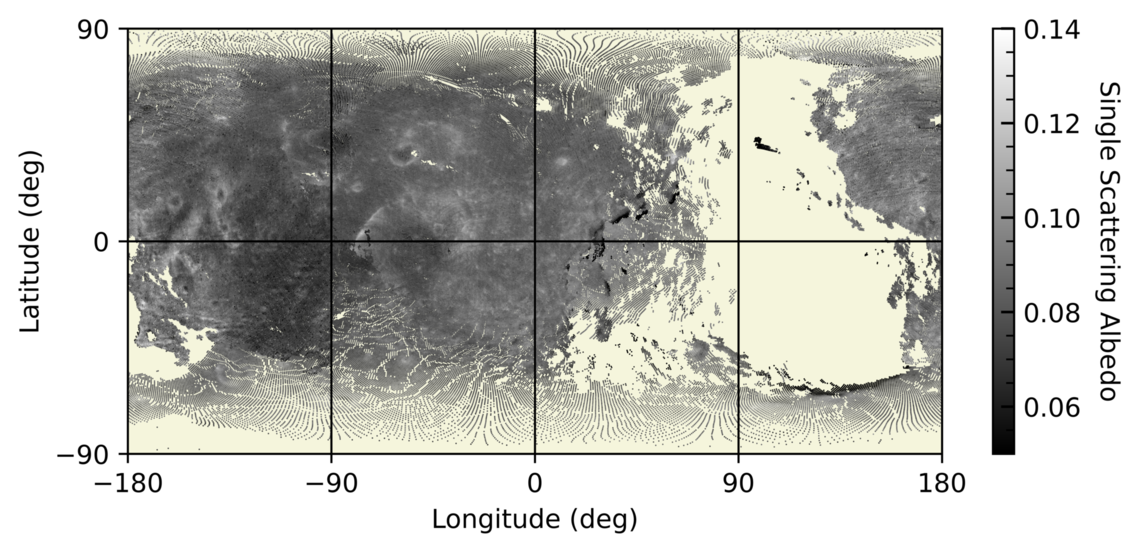}
    \caption{Single-scattering albedo maps in the blue (top), red (center), and IR (bottom) filters.}
    \label{maps_brIR}
\end{figure*}

\begin{figure*}
    \centering
    \includegraphics[width=0.86\textwidth,angle=0]{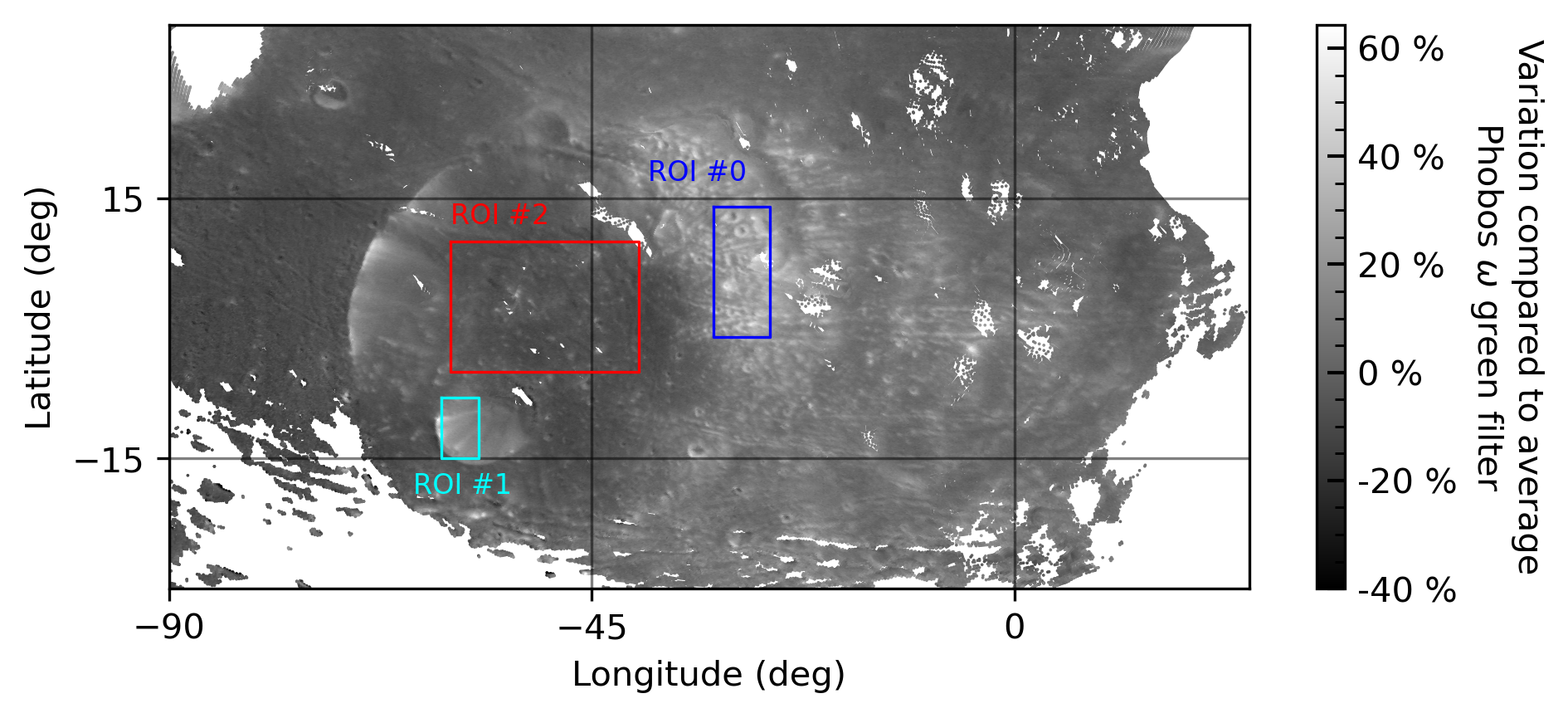}
    \caption{Location of the three ROIs for which the local resolved photometric properties have been investigated (see Table~\ref{tab:hapke_res_roi}).}
    \label{locationROI}
\end{figure*}

\begin{figure*}
    \centering
    \includegraphics[width=0.86\textwidth,angle=0]{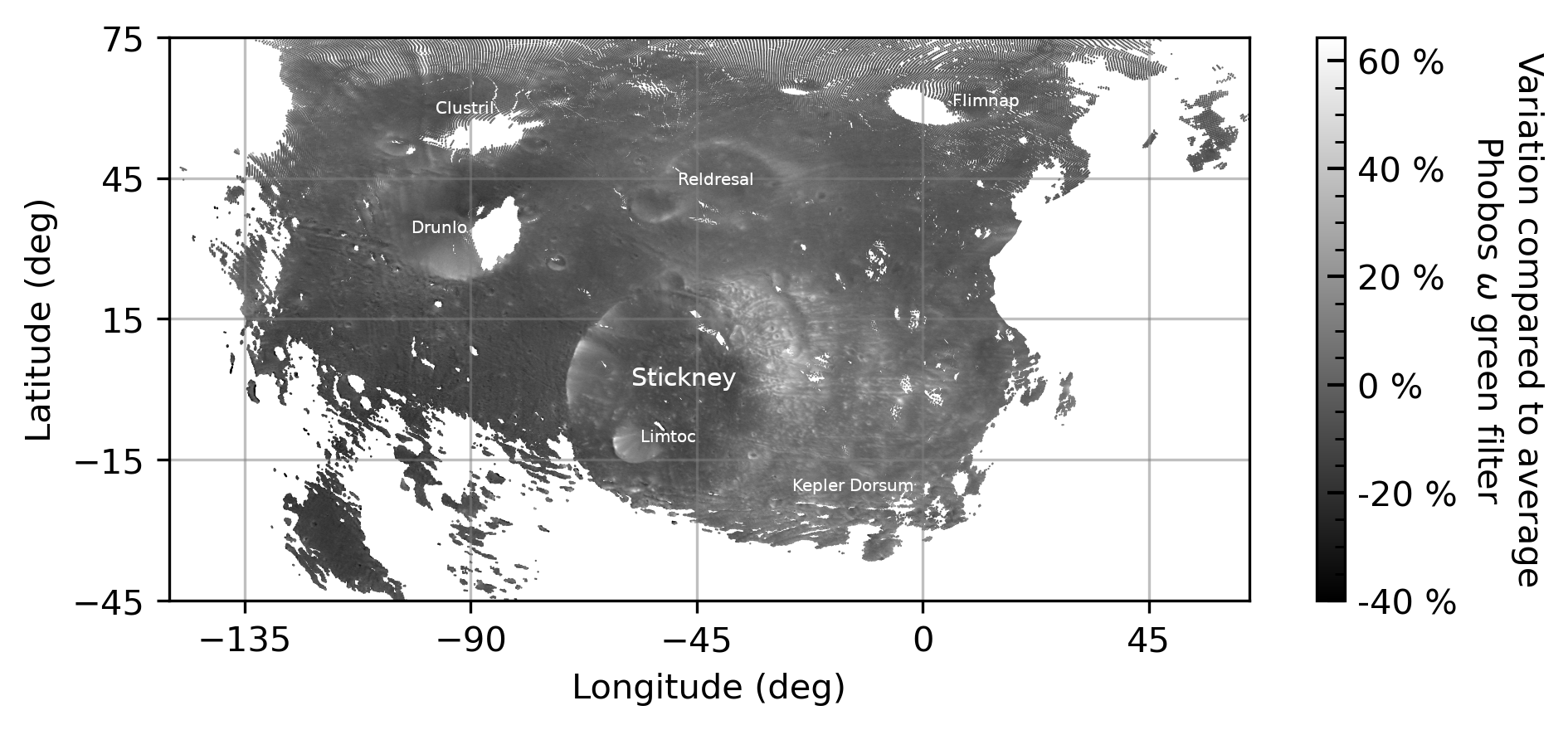} 
    \includegraphics[width=0.86\textwidth,angle=0]{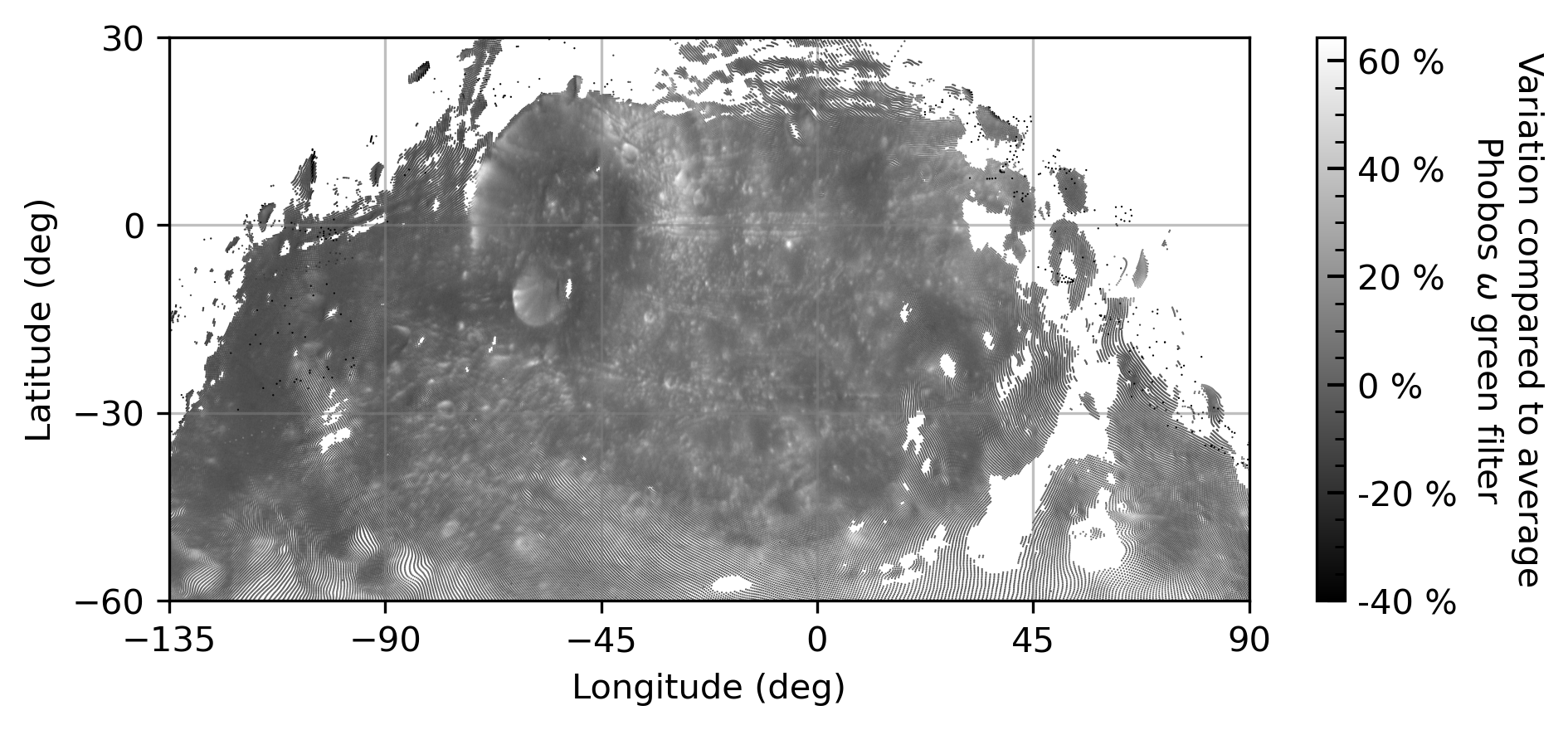}
    \caption{Maps generated from SRC images: Relative variation of the albedo compared to the average value (7.1\%) determined with the green filter and used to normalize the SRC data. Top: From the image 2019-11-17T03:27:01 acquired during K076 orbit. Bottom: From the image 2020-09-26T02:18:40 acquired during L155 orbit. The gray color scale permits better visualization of the albedo-geomorphological feature correlation compared to the colored map presented in Fig. 9.}
    \label{mapsgrey}
\end{figure*}

\clearpage
\twocolumn

\section{Supplementary material: Table}

\begin{table*}[ht]
\centering
\caption{Observing conditions for the blue filter images used in our analysis. Image acquisition times in bold and with an asterisk were utilized to generate the SSA map in the blue filter shown in Fig.~\ref{maps_brIR}.}
\vspace*{\baselineskip}
\begin{tabular}{|ccc|cccc|}
\hline
  Time & Phase ($^{\circ}$) & Res. (m~px$^{-1}$) && Time & Phase ($^{\circ}$) & Res. (m~px$^{-1}$)\\ \hline
2004-05-18T08.33.28.665 &    44.9 &          301 & &  2014-01-01T19.10.05.925 &    51.6 &           65  \\  
2004-07-23T12.38.49.072 &    82.9 &          295 & &  2014-01-08T11.52.06.932 &    36.1 &           61  \\  
2004-08-11T00.29.36.091 &    38.8 &          194 & &  2014-01-11T16.38.18.952 &    70.6 &           44  \\  
2004-12-28T01.18.41.109 &    74.4 &          314 & &  2014-01-15T04.41.34.932 &    26.2 &           50  \\   
2005-04-03T23.50.50.289 &    54.1 &          574 & &  { 2014-01-21T21.20.12.049$^*$} &    25.6 &           71  \\ 
2005-09-16T23.22.24.711 &    64.9 &          306 & &  2014-04-24T20.26.46.220 &    89.1 &           57        \\ 
2005-10-09T22.28.03.232 &    73.6 &          338 & &  2014-05-01T13.13.44.333 &    83.3 &           44        \\ 
2005-12-20T01.13.52.765 &    45.1 &          208 & &  2014-05-25T03.29.26.317 &     3.7 &           45        \\ 
2006-02-19T10.23.26.034 &    59.2 &           90 & &  2014-10-29T00.54.22.958 &    48.2 &           37  \\ 
2006-02-28T16.23.35.892 &    43.3 &           72 & &  2014-11-11T10.30.15.053 &    79.0 &           42        \\ 
2006-03-19T04.19.26.871 &    38.8 &           78 & &  2014-11-14T22.33.16.117 &    12.1 &           44        \\ 
2006-03-30T15.54.56.884 &    58.0 &          106 & &  2014-12-01T20.10.13.085 &     3.7 &           72        \\ 
2006-12-11T21.24.35.587 &    88.8 &           84 & &  2015-02-26T17.45.40.675 &    52.1 &           77  \\
2007-01-01T14.52.00.514 &    50.3 &           50 & &  2015-03-05T10.32.13.602 &    50.9 &           38        \\
2007-01-13T02.25.26.453 &    60.4 &           48 & &  2015-03-15T15.25.01.693 &    21.3 &           53        \\
2007-05-20T15.35.49.397 &    26.1 &           73 & &  2015-12-22T01.29.31.866 &    76.7 &           75        \\
2007-06-01T03.09.40.227 &    16.6 &           50 & &  2016-01-07T23.11.29.820 &    48.7 &           39        \\
2007-06-21T20.40.17.368 &    23.0 &           82 & &  2016-01-18T04.05.04.872 &    13.7 &           51        \\
2007-09-30T11.16.21.173 &    56.6 &           49 & &  2016-05-31T10.10.13.647 &    79.1 &           71        \\
2007-10-21T04.48.38.190 &    58.8 &           51 & &  2016-06-10T15.07.09.627 &    96.8 &           43        \\
2007-10-23T10.27.14.177 &    82.7 &           87 & &  2016-06-14T03.12.14.754 &    56.9 &           56        \\
2007-11-08T16.38.30.167 &    61.9 &           95 & &  2016-06-27T20.15.18.752 &    35.7 &           70        \\
2008-02-18T10.09.28.894 &    95.2 &          108 & &  2016-07-18T06.04.11.869 &    13.2 &           71        \\
2008-03-05T16.24.16.905 &    43.7 &           83 & &  2016-10-27T04.31.36.402 &    82.4 &           48        \\
2008-03-11T02.29.10.017 &    29.0 &          101 & &  2016-10-30T16.37.07.537 &    52.9 &           83  \\
2008-03-24T12.04.54.998 &    66.8 &           66 & &  2017-05-03T23.15.10.413 &    30.5 &           45        \\
2008-04-15T04.21.13.986 &    22.9 &           90 & &  2017-08-26T14.38.37.593 &    67.5 &           77        \\
2008-07-26T01.40.01.680 &    82.1 &          189 & &  2017-09-05T19.34.57.676 &    48.6 &           62        \\
2008-08-03T01.02.11.701 &    45.9 &           52 & &  2017-09-16T00.31.27.740 &    21.7 &           48        \\
2008-08-08T11.10.45.697 &    60.6 &           40 & &  2017-09-29T17.36.25.788 &    10.3 &           86        \\
2009-10-11T20.21.19.853 &    98.2 &           81 & &  2018-01-24T18.26.24.263 &    58.0 &          382        \\
{ 2010-03-16T15.03.11.603$^*$} &    38.2 &           26 & &  2018-02-20T13.16.12.544 &    28.3 &           63        \\
2010-08-17T15.37.03.700 &    68.7 &           33 & &  2018-02-27T05.56.58.541 &    16.3 &          284        \\
2010-09-03T13.19.41.736 &    67.5 &           29 & &  { 2018-03-09T03.19.49.556$^*$} &    37.6 &           25        \\
2011-01-02T21.15.41.907 &    69.3 &           48 & &  2018-03-15T20.00.51.626 &    31.8 &           43        \\
2011-01-06T09.20.01.066 &    56.9 &           96 & &  2018-04-01T10.06.00.910 &    73.1 &           72        \\
2011-01-13T02.11.00.863 &    39.6 &           47 & &  2018-07-29T12.35.52.532 &    40.3 &           40  \\
2011-06-26T15.23.46.135 &    68.3 &           34 & &  2018-08-05T05.18.42.577 &    45.9 &           56        \\
2011-07-06T20.18.16.258 &    39.3 &           82 & &  2018-08-11T22.01.35.567 &    50.8 &           72        \\
{ 2011-07-13T13.07.54.204$^*$} &    39.5 &           30 & &  2018-12-18T02.37.39.119 &    45.6 &           59        \\
2012-05-04T15.19.06.818 &    73.7 &           22 & &  2018-12-24T19.19.20.133 &    35.2 &           62        \\
2012-08-29T19.49.41.150 &    90.6 &           33 & &  2019-01-03T16.43.03.204 &    36.7 &           21        \\
2012-09-09T00.39.41.301 &    44.4 &           60 & &  2019-05-02T19.07.46.663 &    70.5 &          118        \\
2012-09-15T17.26.25.190 &    18.8 &           37 & &  2019-05-09T11.50.11.517 &    50.9 &           63        \\
2012-09-29T02.59.13.303 &    68.4 &           48 & &  2019-11-27T01.01.57.540 &    47.6 &           98        \\
2013-02-25T23.55.01.802 &    61.3 &           66 & &  { 2020-03-28T07.58.23.673$^*$} &    17.6 &           35        \\
2013-03-04T16.40.52.662 &    47.6 &           48 & &  2020-07-27T19.56.16.561 &    65.8 &          174        \\
2013-03-11T09.26.49.782 &    24.2 &           39 & &  2020-08-03T12.38.55.512 &    60.7 &          148        \\
2013-06-22T13.18.05.099 &    84.3 &           53 & &  2020-08-30T07.26.18.560 &    16.8 &           68  \\
2013-06-29T06.02.45.150 &    85.5 &           67 & &  2020-09-06T00.07.35.546 &    16.4 &           79  \\
2013-07-16T03.36.36.295 &    17.6 &           51 & &  2021-07-07T01.41.52.338 &    12.4 &           71  \\
2013-07-22T20.24.17.245 &    29.9 &           40 & &  2021-07-07T01.41.52.338 &    12.4 &           71  \\
2013-07-29T13.07.38.196 &    45.3 &           44 & &  2021-07-07T01.41.52.338 &    12.4 &           71  \\
2013-08-05T05.55.30.359 &    76.1 &           58 & &                          &         &               \\ \hline
\end{tabular}
\label{tab_blue}
\end{table*}

\begin{table*}
\centering
\caption{Observing conditions for the green filter images used in our analysis. Image acquisition times in bold and with an asterisk were utilized to generate the SSA map in the green filter shown in Fig.~\ref{maps}.}
\vspace*{\baselineskip}
\begin{tabular}{|ccc|cccc|}
\hline
  Time & Phase ($^{\circ}$) & Res. (m~px$^{-1}$) && Time & Phase ($^{\circ}$) & Res. (m~px$^{-1}$)\\ \hline
2004-05-18T08.34.37.863 &    39.7   &        301  &   &  2014-01-11T16.37.32.918 &    68.3 &           45   \\ 
2004-07-23T12.40.55.174 &    85.1   &        295  &   &  2014-01-15T04.42.21.884 &    31.4 &           51   \\ 
2004-08-01T18.35.20.745 &    66.3   &        234  &   &  2014-01-21T21.21.30.982 &    18.8 &           65   \\ 
2004-08-11T00.30.26.098 &    42.1   &        194  &   &  2014-01-28T14.08.22.989 &     6.8 &           66   \\ 
2004-08-20T06.25.52.546 &    26.7   &        198  &   &  2014-05-01T13.14.40.242 &    81.6 &           48   \\ 
{ 2005-12-08T13.44.07.867$^*$} &    39.8   &         82  &   &  2014-05-25T03.30.17.325 &    10.1 &           46   \\ 
2006-02-19T10.25.06.831 &    60.5   &         84  &   &  2014-10-29T00.53.44.964 &    46.3 &           38  \\  
2006-02-28T16.24.50.845 &    46.5   &         70  &   &  2014-11-14T22.32.33.097 &     7.3 &           45  \\  
2006-03-05T11.06.39.080 &    46.7   &        167  &   &  2014-12-01T20.09.06.070 &     3.1 &           74  \\  
2006-03-19T04.17.59.890 &    32.2   &         78  &   &  2015-03-05T10.31.32.543 &    50.6 &           38  \\  
2006-03-30T15.53.57.858 &    52.9   &        105  &   &  2015-03-15T15.24.03.604 &    16.9 &           52  \\  
2006-12-11T21.22.58.662 &    85.0   &         79  &   &  2015-08-09T15.37.21.298 &    93.7 &           42  \\  
2007-01-01T14.51.04.620 &    43.7   &         49  &   &  2015-12-22T01.30.48.902 &    74.6 &           75  \\  
2007-01-31T14.18.07.564 &    70.9   &         97  &   &  2016-01-07T23.12.10.844 &    46.6 &           39  \\  
2007-05-20T15.37.00.329 &    24.2   &         74  &   &  2016-01-18T04.06.00.936 &    12.4 &           52  \\  
{ 2007-06-01T03.10.30.291$^*$} &     9.9   &         51  &   &  2016-02-04T01.47.09.956 &    60.2 &           49  \\  
2007-06-21T20.39.00.227 &    28.1   &         79  &   &  2016-02-07T13.55.00.087 &    34.8 &           73  \\  
2007-09-18T23.41.38.182 &    74.1   &        165  &   &  2020-09-06T00.08.48.502 &    10.5 &           79  \\  
2007-09-30T11.17.13.091 &    61.5   &         49  &   &  2016-02-14T06.44.28.089 &    83.0 &           72   \\ 
2007-10-21T04.47.41.066 &    52.3   &         51  &   &  2016-05-31T10.08.54.732 &    76.0 &           73   \\ 
2007-11-08T16.36.36.179 &    56.2   &         98  &   &  2016-06-10T15.06.22.676 &    93.6 &           44   \\ 
2008-03-05T16.23.35.055 &    45.4   &         84  &   &  2016-06-14T03.11.13.688 &    55.3 &           58   \\ 
2008-03-11T02.28.21.031 &    31.8   &        101  &   &  2016-06-27T20.14.12.724 &    33.1 &           70   \\ 
2008-04-15T04.22.37.995 &    17.7   &         94  &   &  2016-07-14T18.00.24.783 &    31.4 &           34   \\ 
2008-07-26T01.38.11.149 &    77.2   &        178  &   &  2016-07-18T06.03.06.831 &     9.7 &           71  \\  
2008-08-08T11.11.29.732 &    65.3   &         41  &   &  2016-10-27T04.30.48.391 &    84.4 &           48  \\  
2008-08-30T03.37.37.133 &    66.9   &        189  &   &  2016-10-30T16.35.39.540 &    54.3 &           83  \\  
2009-05-26T18.32.14.591 &    81.7   &         76  &   &  2016-11-20T02.28.46.441 &     6.7 &          112  \\  
2009-10-11T20.20.27.727 &    96.5   &         87  &   &  2017-05-03T23.15.55.366 &    29.8 &           45  \\  
{ 2010-03-16T15.03.39.612$^*$} &    42.1   &         26  &   &  2017-08-26T14.39.57.635 &    66.3 &           77  \\  
2010-03-19T19.36.10.117 &    45.8   &        272  &   &  2017-09-05T19.36.02.769 &    47.5 &           62  \\  
2010-04-05T17.09.25.627 &    32.0   &         78  &   &  2017-09-16T00.32.19.693 &    20.4 &           48  \\  
2010-08-17T15.37.36.714 &    71.8   &         33  &   &  2017-10-09T22.32.03.743 &    30.5 &          168  \\  
2010-09-03T13.19.07.787 &    65.8   &         31  &   &  2018-01-24T18.25.53.544 &    57.1 &          382  \\  
2011-01-06T09.21.45.999 &    58.9   &         94  &   &  2018-02-20T13.15.11.524 &    26.4 &           63  \\  
2011-01-13T02.11.50.968 &    41.8   &         46  &   &  { 2018-03-09T03.20.35.526$^*$} &    39.0 &           25  \\  
2011-06-26T15.23.10.201 &    68.2   &         35  &   &  2018-03-12T15.17.45.693 &    10.1 &           96   \\ 
2011-07-06T20.17.37.337 &    39.7   &         83  &   &  2018-03-15T20.02.06.579 &    31.8 &           43   \\ 
{ 2011-07-13T13.08.24.203$^*$} &    40.1   &         30  &   &  2018-12-18T02.38.38.133 &    44.9 &           59   \\ 
2012-04-27T22.31.42.798 &    51.7   &         15  &   &  2018-12-24T19.20.20.281 &    33.4 &           62   \\ 
2012-05-04T15.18.44.825 &    70.6   &         22  &   &  2019-01-03T16.42.49.246 &    38.3 &           21   \\ 
{ 2012-05-14T20.08.33.912$^*$} &    15.1   &         42  &   &  2019-01-07T04.42.13.402 &    24.4 &          334   \\ 
2012-09-15T17.27.06.215 &    20.4   &         37  &   &  2019-01-10T09.24.28.312 &    30.6 &           27  \\  
2012-10-02T15.04.06.324 &    21.2   &         66  &   &  2020-09-06T00.08.48.502 &    10.5 &           79  \\  
2012-10-16T00.35.32.498 &    64.2   &         89  &   &  2020-09-06T00.08.48.502 &    10.5 &           79  \\  
2013-02-25T23.54.01.692 &    59.7   &         66  &   &  2020-09-06T00.08.48.502 &    10.5 &           79  \\  
2013-03-01T04.38.23.691 &    94.7   &         18  &   &  2019-06-01T18.39.05.410 &    45.0 &           36  \\  
2013-03-04T16.40.07.689 &    45.8   &         48  &   &  2019-11-06T22.48.04.505 &    74.0 &           27  \\  
2013-03-11T09.26.12.743 &    20.8   &         39  &   &  2019-11-13T15.29.57.622 &    46.6 &           48  \\  
2013-06-22T13.17.11.055 &    83.8   &         54  &   &  2020-03-21T15.16.21.663 &    13.5 &           36  \\  
2013-06-29T06.02.08.211 &    84.9   &         70  &   &  2020-03-28T07.59.02.767 &    18.5 &           35  \\  
2013-07-16T03.35.37.269 &    15.4   &         52  &   &  2020-07-27T19.57.49.697 &    68.5 &          174  \\  
2013-07-29T13.08.26.292 &    48.3   &         43  &   &  { 2020-08-03T12.40.13.598$^*$} &    62.9 &          148  \\  
2013-08-05T05.56.34.342 &    79.5   &         59  &   &  2020-09-06T00.08.48.502 &    10.5 &           79  \\  
2013-12-09T04.48.47.777 &    79.8   &         71  &   &  2021-01-19T16.47.07.883 &    20.9 &           30  \\
2013-12-15T21.35.25.808 &    81.3   &         46  &   &  2021-02-02T02.12.56.920 &    40.5 &           46  \\
2014-01-01T19.11.05.887 &    53.9   &         66  &   &  2021-02-15T11.38.55.052 &    50.7 &           79  \\
2014-01-04T23.53.37.815 &    71.2   &         22  &   &  2021-07-07T01.40.47.246 &    14.5 &           71  \\
2014-01-08T11.53.05.971 &    35.6   &         58  &   &                          &         &               \\ \hline
\end{tabular}
\label{tab_green}
\end{table*}

\begin{table*}[ht]
\centering
\caption{Observing conditions for the red filter images used in our analysis. Image acquisition times in bold and with an asterisk were utilized to generate the SSA map in the red filter shown in Fig.~\ref{maps_brIR}.}
\vspace*{\baselineskip}
\begin{tabular}{|ccc|cccc|}
\hline
  Time & Phase ($^{\circ}$) & Res. (m~px$^{-1}$) && Time & Phase ($^{\circ}$) & Res. (m~px$^{-1}$)\\ \hline
2004-05-18T08.31.15.476 &    55.1  &         314& &  2015-02-26T17.48.26.517 &    54.6 &           82  \\
2004-08-01T18.32.27.492 &    61.7  &         244& &  2015-03-05T10.33.33.517 &    53.1 &           40  \\
2004-08-11T00.27.58.276 &    34.8  &         202& &  { 2015-03-15T15.26.57.553$^*$} &    31.8 &           56  \\
2004-08-20T06.23.13.451 &     7.4  &         207& &  2015-04-01T13.03.27.000 &    45.5 &           52  \\
2005-09-16T23.17.27.117 &    70.5  &         317& &  2015-08-09T15.35.00.271 &    86.8 &           47  \\
{ 2006-02-28T16.20.59.843$^*$} &    39.2  &          78& &  2015-12-22T01.27.00.731 &    81.2 &           79  \\
2007-01-13T02.23.44.450 &    52.5  &          49& &  2015-12-28T18.18.33.000 &    89.8 &           40  \\
2007-01-31T14.12.33.656 &    56.3  &         101& &  2016-01-07T23.10.09.784 &    54.3 &           40  \\
2007-05-20T15.33.38.282 &    32.9  &          73& &  2016-01-18T04.03.16.914 &    22.2 &           53  \\
{ 2007-06-01T03.08.09.326$^*$} &    29.1  &          51& &  2021-07-07T01.43.58.330 &    15.7 &           74  \\
2007-09-30T11.14.41.116 &    48.3  &          51& &  2021-07-07T01.43.58.330 &    15.7 &           74  \\
2007-10-23T10.28.43.262 &    91.5  &          91& &  2016-05-31T10.12.35.566 &    85.2 &           71  \\
2007-10-30T10.45.31.160 &    67.6  &         149& &  2016-06-14T03.13.59.747 &    61.1 &           54  \\
2021-07-07T01.43.58.330 &    15.7  &          74& &  2016-06-27T20.17.26.828 &    42.6 &           73  \\
2008-03-05T16.25.36.944 &    42.8  &          86& &  2016-07-14T17.58.45.810 &    26.6 &           36  \\
2008-03-29T22.06.24.857 &    54.5  &          86& &  2016-07-18T06.06.15.792 &    23.6 &           74  \\
2008-04-15T04.18.45.936 &    34.1  &          88& &  2016-10-27T04.33.10.439 &    78.8 &           50  \\
2008-08-08T11.09.27.733 &    52.4  &          39& &  2016-10-30T16.39.57.536 &    51.7 &           86  \\
2008-08-30T03.32.40.945 &    55.0  &         196& &  2017-05-03T23.13.44.383 &    35.1 &           47  \\
2009-06-13T07.27.05.902 &    65.2  &         141& &  2017-08-26T14.36.01.674 &    70.6 &           80  \\
{ 2010-03-16T15.02.18.589$^*$} &    33.1  &          27& &  2017-09-05T19.32.50.700 &    52.3 &           64  \\
2010-03-26T12.16.38.870 &    29.8  &         104& &  2017-09-16T00.29.46.791 &    28.5 &           50  \\
2010-09-13T18.15.22.726 &    53.1  &          45& &  2017-09-29T17.39.34.858 &    22.3 &           89  \\
2011-01-06T09.16.26.898 &    54.3  &         103& &  { 2018-02-20T13.18.11.558$^*$} &    35.0 &           65  \\
2011-01-13T02.09.19.969 &    37.9  &          50& &  2018-02-27T05.57.49.617 &    18.3 &          286  \\
2011-07-07T01.43.58.330 &    15.7  &          74& &  2018-03-12T15.21.54.714 &    12.3 &          100  \\
2011-07-06T20.19.31.304 &    41.1  &          85& &  2018-07-29T12.34.29.476 &    44.1 &           42  \\
{ 2011-07-13T13.06.57.174$^*$} &    41.0  &          31& &  2018-11-28T00.30.29.132 &    70.4 &          334  \\
2012-04-27T22.30.56.761 &    42.4  &          16& &  2018-12-18T02.35.43.199 &    48.9 &           62  \\
2012-05-04T15.19.49.833 &    80.0  &          23& &  2018-12-24T19.17.25.186 &    41.1 &           64  \\
2012-08-29T19.48.33.200 &    92.3  &          34& &  2019-01-07T04.40.48.330 &    30.0 &          336  \\
2012-09-29T03.00.53.306 &    70.8  &          50& &  2019-01-10T09.25.48.301 &    26.9 &           29  \\
2012-10-02T15.07.48.316 &    36.6  &          69& &  2019-05-02T19.05.50.622 &    71.0 &          122  \\
2021-07-07T01.43.58.330 &    15.7  &          74& &  2019-05-09T11.48.03.396 &    51.5 &           65  \\
2013-03-01T04.37.28.654 &    87.3  &          19& &  2019-05-16T04.30.33.622 &    34.5 &          143  \\
2013-03-04T16.42.20.763 &    52.7  &          50& &  2019-06-01T18.40.57.431 &    50.1 &           37  \\
2013-03-11T09.28.00.766 &    33.1  &          41& &  2019-11-13T15.39.08.561 &    73.4 &          146  \\
2013-06-22T13.19.47.100 &    85.3  &          54& &  2020-03-21T15.15.24.729 &    26.0 &           37  \\
2013-07-16T03.38.20.154 &    26.2  &          50& &  { 2020-03-28T07.57.10.766$^*$} &    24.3 &           36  \\
2013-07-22T20.22.54.244 &    18.9  &          40& &  2020-07-27T19.53.18.343 &    61.4 &          181  \\
2013-08-05T05.53.33.290 &    70.0  &          58& &  2021-07-07T01.43.58.330 &    15.7 &           74  \\
2014-01-01T19.08.10.915 &    48.9  &          67& &  2020-08-23T14.43.11.463 &    31.0 &           62  \\
2014-01-11T16.39.40.861 &    75.7  &          43& &  2020-08-30T07.24.14.519 &    27.7 &           71  \\
2014-05-25T03.27.51.351 &     9.5  &          46& &  2020-09-06T00.05.16.637 &    30.2 &           82  \\
2014-10-29T00.55.32.019 &    53.4  &          37& &  2020-09-12T16.45.43.688 &    35.1 &           97  \\
2014-11-14T22.34.35.026 &    23.8  &          44& &  2021-02-05T14.23.30.898 &    46.7 &          102  \\
2014-12-01T20.12.15.065 &    16.1  &          73& &  2021-07-07T01.43.58.330 &    15.7 &           74  \\ \hline
\end{tabular}
\label{tab_red}
\end{table*}

\begin{table*}[ht]
\centering
\caption{Observing conditions for the IR filter images used in our analysis. Image acquisition times in bold and with an asterisk were utilized to generate the SSA map in the IR filter shown in Fig.~\ref{maps_brIR}.}
\vspace*{\baselineskip}
\begin{tabular}{|ccc|cccc|}
\hline
  Time & Phase ($^{\circ}$) & Res. (m~px$^{-1}$) && Time & Phase ($^{\circ}$) & Res. (m~px$^{-1}$)\\ \hline
2004-05-18T08.36.49.563 &    31.6  &         313 & &  2013-06-22T13.15.21.036 &    83.2 &           57  \\
2004-08-01T18.37.14.321 &    70.7  &         243 & &  2013-07-16T03.33.30.186 &    18.3 &           59  \\
2004-08-11T00.32.04.462 &    49.9  &         202 & &  2013-08-05T05.58.48.285 &    86.2 &           65  \\
2004-08-20T06.27.36.457 &    39.2  &         206 & &  2014-01-01T19.13.02.863 &    59.4 &           69  \\
2004-12-28T01.14.47.902 &    81.8  &         327 & &  2014-01-04T23.52.54.833 &    66.7 &           23  \\
2005-04-03T23.45.17.358 &    68.1  &         600 & &  2014-01-15T04.43.57.863 &    42.2 &           55  \\
2005-09-16T23.29.58.268 &    58.7  &         317 & &  2014-01-28T14.06.15.028 &    15.7 &           72  \\
2005-10-09T22.32.10.211 &    63.4  &         351 & &  2014-02-04T06.53.16.916 &    12.7 &           76  \\
{ 2005-12-08T13.46.34.824$^*$} &    27.8  &          87 & &  2014-04-24T20.29.52.170 &    83.1 &           64  \\
2005-12-20T01.18.41.671 &    29.4  &         238 & &  2014-05-25T03.31.59.342 &    22.5 &           49  \\
2006-02-28T16.27.07.849 &    53.9  &          70 & &  2014-05-31T20.13.13.313 &    31.5 &           42  \\
2006-03-05T11.12.52.756 &    59.0  &         177 & &  2014-10-29T00.52.22.993 &    44.6 &           42   \\
{ 2007-01-01T14.49.22.552$^*$} &    31.5  &          48 & &  2014-11-11T10.32.20.109 &    87.4 &           42         \\
2007-01-13T02.28.09.510 &    74.1  &          52 & &  2015-03-05T10.30.16.591 &    51.8 &           39         \\
2007-01-31T14.21.48.420 &    80.8  &         102 & &  2015-12-22T01.33.16.888 &    71.1 &           78         \\
2007-05-20T15.39.27.272 &    25.1  &          81 & &  2016-01-07T23.13.29.874 &    44.5 &           40         \\
2007-06-21T20.36.45.177 &    38.9  &          77 & &  2016-01-18T04.07.53.836 &    18.3 &           55         \\
2007-07-14T19.41.30.169 &    68.2  &         232 & &  2016-02-04T01.45.26.926 &    61.4 &           51         \\
2007-09-18T23.44.21.375 &    82.6  &         169 & &  2016-02-07T13.52.19.993 &    35.1 &           75         \\
2007-09-30T11.18.51.049 &    71.2  &          50 & &  2021-07-07T01.38.41.255 &    23.6 &           74         \\
2007-10-21T04.45.55.065 &    40.6  &          52 & &  2016-02-14T06.41.59.112 &    82.3 &           73         \\
2007-10-23T10.25.01.046 &    69.3  &          89 & &  2016-06-14T03.09.01.740 &    53.8 &           65         \\
2008-02-18T10.06.42.123 &    93.9  &         111 & &  2016-06-27T20.12.04.830 &    31.3 &           73  \\
2008-03-05T16.22.12.043 &    50.4  &          88 & &  2016-07-14T18.01.29.791 &    39.2 &           36  \\
2008-03-11T02.26.45.987 &    39.6  &         105 & &  2016-07-18T06.01.01.853 &    13.6 &           74  \\
2008-03-29T22.09.46.065 &    43.6  &          90 & &  2016-10-27T04.29.14.353 &    88.5 &           50  \\
2008-04-15T04.25.40.016 &    12.3  &         105 & &  2016-10-30T16.32.50.500 &    58.4 &           86  \\
2008-08-03T01.03.30.663 &    57.3  &          53 & &  2017-08-26T14.42.33.773 &    64.9 &           80  \\
2008-08-08T11.13.05.744 &    74.6  &          46 & &  2017-09-05T19.38.09.746 &    47.4 &           64  \\
2008-08-30T03.40.51.662 &    75.4  &         197 & &  { 2017-09-16T00.34.00.783$^*$} &    23.3 &           50  \\
2009-06-13T07.20.58.625 &    40.6  &         148 & &  2017-09-29T17.31.36.673 &    11.0 &           89  \\
2010-03-16T15.04.33.603 &    50.7  &          27 & &  2018-02-20T13.13.11.502 &    26.9 &           65  \\
2010-03-26T12.21.21.753 &    57.9  &         116 & &  2018-02-27T05.55.41.633 &    16.0 &          286  \\
2010-04-05T17.12.06.769 &    41.3  &          76 & &  2018-03-12T15.14.58.670 &    21.7 &          100  \\
2010-08-17T15.38.41.748 &    78.3  &          34 & &  2018-11-28T00.32.50.185 &    72.2 &          335  \\
2010-08-27T20.33.05.658 &    48.0  &          34 & &  2018-12-18T02.40.33.242 &    45.7 &           62  \\
2011-01-02T21.18.06.981 &    73.0  &          48 & &  2018-12-24T19.22.16.201 &    33.2 &           64  \\
2011-01-06T09.25.00.958 &    63.9  &          95 & &  2019-01-03T16.42.19.246 &    45.8 &           22  \\
2011-01-13T02.13.24.899 &    47.8  &          47 & &  2019-01-10T09.23.35.254 &    38.1 &           28  \\
2011-07-06T20.16.19.281 &    43.0  &          87 & &  2019-05-02T19.10.43.657 &    71.6 &          122  \\
2011-07-13T13.09.21.233 &    43.4  &          31 & &  2019-05-09T11.53.25.402 &    54.2 &           65  \\
2012-04-21T05.46.23.841 &    79.8  &          45 & &  2019-05-16T04.36.44.608 &    41.0 &          143  \\
2012-04-27T22.32.12.778 &    59.8  &          15 & &  2019-06-01T18.37.51.480 &    45.0 &           37  \\
{ 2012-05-04T15.18.01.809$^*$} &    65.3  &          23 & &  2019-06-01T18.37.51.480 &    45.0 &           37  \\
2012-05-14T20.07.15.886 &     3.8  &          44 & &  2019-11-06T22.48.21.532 &    75.1 &           29  \\
2012-08-29T19.51.24.174 &    88.0  &          34 & &  2019-11-13T15.19.57.654 &    78.9 &          144  \\
2012-09-09T00.42.57.235 &    48.1  &          63 & &  2020-03-21T15.16.59.756 &    19.7 &           37  \\
2012-09-15T17.28.25.245 &    27.6  &          39 & &  2020-03-28T08.00.15.675 &    25.2 &           36  \\
2012-09-29T02.56.40.324 &    66.8  &          50 & &  2020-07-27T20.00.49.425 &    74.2 &          181  \\
{ 2012-10-02T15.01.40.309$^*$} &    16.3  &          69 & &  2020-08-03T12.42.42.633 &    67.8 &          154  \\
2012-10-16T00.32.19.438 &    60.5  &          93 & &  2020-08-23T14.47.56.399 &    29.2 &           62  \\
2013-02-25T23.52.04.765 &    57.8  &          68 & &  { 2020-08-30T07.29.27.478$^*$} &    16.3 &           71  \\
2013-03-01T04.38.58.683 &    99.2  &          19 & &  2021-01-19T16.46.19.892 &    20.1 &           32  \\
2013-03-04T16.38.40.694 &    44.3  &          50 & &  2021-02-15T11.36.05.072 &    44.7 &           82  \\
2013-03-11T09.25.01.759 &    19.7  &          41 & &  2021-07-07T01.38.41.255 &    23.6 &           74  \\ \hline
\end{tabular}
\label{tab_ir}
\end{table*}

\clearpage
\onecolumn

\begin{center}
\begin{longtable}{|cccc|}
\caption{Observing conditions for the SRC camera images used in our analysis of the Phobos opposition surge.}
\label{tab_SRC} \\
  \hline
  \multicolumn{1}{|c}{Time} & \multicolumn{1}{c}{Phase ($^{\circ}$)} & \multicolumn{1}{c}{Helioc. dist. (au)} & \multicolumn{1}{c|}{Res. (m~px$^{-1}$)}\\
  \hline
\endfirsthead
\hline 
\multicolumn{1}{|c}{Time} & \multicolumn{1}{c}{Phase ($^{\circ}$)} & \multicolumn{1}{c}{Helioc. dist. (au)} & \multicolumn{1}{c|}{Res. (m~px$^{-1}$)}\\ \hline 
\endhead
\hline
\multicolumn{4}{c}{\tablename\ \thetable\ -- \textit{Continued on next page}} \\ 
\endfoot
\hline \hline
\endlastfoot
2019-11-17T03:27:01.647 & 17.13 & 1.639 & 21.86 \\ 
2019-11-17T03:27:12.547 & 16.27 & 1.639 & 21.83 \\ 
2019-11-17T03:27:23.448 & 15.41 & 1.639 & 21.81 \\ 
2019-11-17T03:27:34.348 & 14.55 & 1.639 & 21.79 \\ 
2019-11-17T03:27:45.248 & 13.69 & 1.639 & 21.78 \\ 
2019-11-17T03:27:56.149 & 12.83 & 1.639 & 21.77 \\ 
2019-11-17T03:28:07.049 & 11.96 & 1.639 & 21.77 \\ 
2019-11-17T03:28:17.949 & 11.10 & 1.639 & 21.77 \\ 
2019-11-17T03:28:28.850 & 10.24 & 1.639 & 21.78 \\ 
2019-11-17T03:28:39.750 & 9.38 & 1.639 & 21.79 \\ 
2019-11-17T03:28:50.651 & 8.52 & 1.639 & 21.81 \\ 
2019-11-17T03:29:01.551 & 7.66 & 1.639 & 21.83 \\ 
2019-11-17T03:29:12.451 & 6.80 & 1.639 & 21.86 \\ 
2019-11-17T03:29:23.352 & 5.95 & 1.639 & 21.90 \\ 
2019-11-17T03:29:34.252 & 5.11 & 1.639 & 21.93 \\ 
2019-11-17T03:29:45.153 & 4.27 & 1.639 & 21.98 \\ 
2019-11-17T03:29:56.053 & 3.44 & 1.639 & 22.02 \\ 
2019-11-17T03:30:06.953 & 2.63 & 1.639 & 22.08 \\ 
2019-11-17T03:30:17.854 & 1.87 & 1.639 & 22.13 \\ 
2019-11-17T03:30:28.754 & 1.21 & 1.639 & 22.20 \\ 
2019-11-17T03:30:39.655 & 0.92 & 1.639 & 22.26 \\ 
2019-11-17T03:30:50.555 & 1.28 & 1.639 & 22.34 \\ 
2019-11-17T03:31:01.456 & 1.94 & 1.639 & 22.41 \\ 
2019-11-17T03:31:12.356 & 2.69 & 1.639 & 22.49 \\ 
2019-11-17T03:31:23.256 & 3.47 & 1.639 & 22.58 \\ 
2019-11-17T03:31:34.157 & 4.25 & 1.639 & 22.67 \\ 
2019-11-17T03:31:45.057 & 5.04 & 1.639 & 22.77 \\ 
2019-11-17T03:31:55.958 & 5.82 & 1.639 & 22.87 \\ 
2019-11-17T03:32:06.858 & 6.60 & 1.639 & 22.97 \\ 
2019-11-17T03:32:17.759 & 7.38 & 1.639 & 23.08 \\ 
2019-11-17T03:32:28.659 & 8.15 & 1.639 & 23.19 \\ 
2019-11-17T03:32:39.559 & 8.91 & 1.639 & 23.31 \\ 
2019-11-17T03:32:50.460 & 9.66 & 1.639 & 23.43 \\ 
2019-11-17T03:33:01.360 & 10.41 & 1.639 & 23.56 \\ 
2019-11-17T03:33:12.261 & 11.16 & 1.639 & 23.69 \\ 
2019-11-17T03:33:23.161 & 11.89 & 1.639 & 23.83 \\ 
2019-11-17T03:33:34.062 & 12.62 & 1.639 & 23.96 \\ 
2019-11-17T03:33:44.962 & 13.34 & 1.639 & 24.11 \\ 
2019-11-17T03:33:55.863 & 14.05 & 1.639 & 24.25 \\ 
2019-11-17T03:34:06.763 & 14.75 & 1.639 & 24.40 \\ 
2019-11-30T12:52:22.472 & 6.89 & 1.628 & 30.44 \\ 
2019-11-30T12:52:24.653 & 6.76 & 1.628 & 30.46 \\ 
2019-11-30T12:52:26.833 & 6.63 & 1.628 & 30.48 \\ 
2019-11-30T12:52:29.013 & 6.50 & 1.628 & 30.50 \\ 
2019-11-30T12:52:31.193 & 6.37 & 1.628 & 30.52 \\ 
2019-11-30T12:52:33.373 & 6.24 & 1.628 & 30.54 \\ 
2019-11-30T12:52:35.553 & 6.11 & 1.628 & 30.56 \\ 
2019-11-30T12:52:37.733 & 5.99 & 1.628 & 30.57 \\ 
2019-11-30T12:52:39.913 & 5.86 & 1.628 & 30.59 \\ 
2019-11-30T12:52:42.093 & 5.73 & 1.628 & 30.61 \\ 
2019-11-30T12:52:44.273 & 5.60 & 1.628 & 30.63 \\ 
2019-11-30T12:52:46.453 & 5.47 & 1.628 & 30.65 \\ 
2019-11-30T12:52:48.633 & 5.34 & 1.628 & 30.67 \\ 
2019-11-30T12:52:50.813 & 5.21 & 1.628 & 30.69 \\ 
2019-11-30T12:52:52.994 & 5.09 & 1.628 & 30.71 \\ 
2019-11-30T12:52:55.174 & 4.96 & 1.628 & 30.73 \\ 
2019-11-30T12:52:57.354 & 4.83 & 1.628 & 30.75 \\ 
2019-11-30T12:52:59.534 & 4.70 & 1.628 & 30.77 \\ 
2019-11-30T12:53:01.714 & 4.58 & 1.628 & 30.79 \\ 
2019-11-30T12:53:03.894 & 4.45 & 1.628 & 30.81 \\ 
2019-11-30T12:53:06.074 & 4.32 & 1.628 & 30.83 \\ 
2019-11-30T12:53:08.254 & 4.19 & 1.628 & 30.85 \\ 
2019-11-30T12:53:10.434 & 4.07 & 1.628 & 30.88 \\ 
2019-11-30T12:53:12.614 & 3.94 & 1.628 & 30.90 \\ 
2019-11-30T12:53:14.794 & 3.81 & 1.628 & 30.92 \\ 
2019-11-30T12:53:16.974 & 3.69 & 1.628 & 30.94 \\ 
2019-11-30T12:53:19.154 & 3.56 & 1.628 & 30.96 \\ 
2019-11-30T12:53:21.335 & 3.44 & 1.628 & 30.99 \\ 
2019-11-30T12:53:23.515 & 3.31 & 1.628 & 31.01 \\ 
2019-11-30T12:53:25.695 & 3.19 & 1.628 & 31.03 \\ 
2019-11-30T12:53:27.875 & 3.06 & 1.628 & 31.05 \\ 
2019-11-30T12:53:30.055 & 2.94 & 1.628 & 31.08 \\ 
2019-11-30T12:53:32.235 & 2.81 & 1.628 & 31.10 \\ 
2019-11-30T12:53:34.415 & 2.69 & 1.628 & 31.12 \\ 
2019-11-30T12:53:36.595 & 2.56 & 1.628 & 31.15 \\ 
2019-11-30T12:53:38.775 & 2.44 & 1.628 & 31.17 \\ 
2019-11-30T12:53:40.955 & 2.32 & 1.628 & 31.19 \\ 
2019-11-30T12:53:43.135 & 2.19 & 1.628 & 31.22 \\ 
2019-11-30T12:53:45.315 & 2.07 & 1.628 & 31.24 \\ 
2019-11-30T12:53:47.495 & 1.95 & 1.628 & 31.26 \\ 
2019-11-30T12:53:49.676 & 1.83 & 1.628 & 31.29 \\ 
2019-11-30T12:53:52.401 & 1.68 & 1.628 & 31.32 \\ 
2019-11-30T12:53:54.581 & 1.55 & 1.628 & 31.34 \\ 
2019-11-30T12:53:56.761 & 1.44 & 1.628 & 31.37 \\ 
2019-11-30T12:53:58.941 & 1.32 & 1.628 & 31.39 \\ 
2019-11-30T12:54:01.121 & 1.20 & 1.628 & 31.42 \\ 
2019-11-30T12:54:03.301 & 1.08 & 1.628 & 31.44 \\ 
2019-11-30T12:54:05.481 & 0.96 & 1.628 & 31.47 \\ 
2019-11-30T12:54:07.661 & 0.85 & 1.628 & 31.49 \\ 
2019-11-30T12:54:09.841 & 0.74 & 1.628 & 31.52 \\ 
2019-11-30T12:54:12.021 & 0.64 & 1.628 & 31.55 \\ 
2019-11-30T12:54:13.111 & 0.59 & 1.628 & 31.56 \\ 
2019-11-30T12:54:14.201 & 0.54 & 1.628 & 31.57 \\ 
2019-11-30T12:54:15.291 & 0.50 & 1.628 & 31.58 \\ 
2019-11-30T12:54:16.381 & 0.46 & 1.628 & 31.60 \\ 
2019-11-30T12:54:17.472 & 0.42 & 1.628 & 31.61 \\ 
2019-11-30T12:54:18.562 & 0.39 & 1.628 & 31.62 \\ 
2019-11-30T12:54:19.652 & 0.37 & 1.628 & 31.64 \\ 
2019-11-30T12:54:20.742 & 0.35 & 1.628 & 31.65 \\ 
2019-11-30T12:54:21.832 & 0.35 & 1.628 & 31.66 \\ 
2019-11-30T12:54:22.922 & 0.36 & 1.628 & 31.68 \\ 
2019-11-30T12:54:24.012 & 0.38 & 1.628 & 31.69 \\ 
2019-11-30T12:54:25.102 & 0.40 & 1.628 & 31.70 \\ 
2019-11-30T12:54:26.192 & 0.43 & 1.628 & 31.72 \\ 
2019-11-30T12:54:27.282 & 0.47 & 1.628 & 31.73 \\ 
2019-11-30T12:54:28.372 & 0.52 & 1.628 & 31.74 \\ 
2019-11-30T12:54:29.462 & 0.56 & 1.628 & 31.76 \\ 
2019-11-30T12:54:30.552 & 0.61 & 1.628 & 31.77 \\ 
2019-11-30T12:54:31.642 & 0.66 & 1.628 & 31.78 \\ 
2019-11-30T12:54:32.732 & 0.71 & 1.628 & 31.80 \\ 
2019-11-30T12:54:34.367 & 0.79 & 1.628 & 31.82 \\ 
2019-11-30T12:54:38.727 & 1.01 & 1.628 & 31.87 \\ 
2019-11-30T12:54:43.087 & 1.24 & 1.628 & 31.93 \\ 
2019-11-30T12:54:47.448 & 1.47 & 1.628 & 31.98 \\ 
2019-11-30T12:54:51.808 & 1.70 & 1.628 & 32.04 \\ 
2019-11-30T12:54:56.168 & 1.93 & 1.628 & 32.10 \\ 
2019-11-30T12:55:00.528 & 2.17 & 1.628 & 32.16 \\ 
2019-11-30T12:55:04.888 & 2.40 & 1.628 & 32.22 \\ 
2019-11-30T12:55:09.248 & 2.63 & 1.628 & 32.28 \\ 
2019-11-30T12:55:13.609 & 2.86 & 1.628 & 32.33 \\ 
2019-11-30T12:55:17.969 & 3.10 & 1.628 & 32.40 \\ 
2019-11-30T12:55:22.329 & 3.33 & 1.628 & 32.46 \\ 
2019-11-30T12:55:26.689 & 3.56 & 1.628 & 32.52 \\ 
2019-11-30T12:55:31.049 & 3.79 & 1.628 & 32.58 \\ 
2019-11-30T12:55:35.409 & 4.01 & 1.628 & 32.64 \\ 
2019-11-30T12:55:39.770 & 4.24 & 1.628 & 32.71 \\ 
2019-11-30T12:55:44.130 & 4.47 & 1.628 & 32.77 \\ 
2019-11-30T12:55:48.490 & 4.70 & 1.628 & 32.83 \\ 
2019-11-30T12:55:52.850 & 4.92 & 1.628 & 32.90 \\ 
2019-11-30T12:55:57.210 & 5.14 & 1.628 & 32.96 \\ 
2020-09-19T09:35:08.001 & 2.52 & 1.403 & 26.39 \\ 
2020-09-19T09:35:09.091 & 2.45 & 1.403 & 26.40 \\ 
2020-09-19T09:35:10.181 & 2.38 & 1.403 & 26.42 \\ 
2020-09-19T09:35:11.271 & 2.31 & 1.403 & 26.43 \\ 
2020-09-19T09:35:12.361 & 2.25 & 1.403 & 26.44 \\ 
2020-09-19T09:35:13.451 & 2.17 & 1.403 & 26.45 \\ 
2020-09-19T09:35:14.541 & 2.10 & 1.403 & 26.47 \\ 
2020-09-19T09:35:15.631 & 2.04 & 1.403 & 26.48 \\ 
2020-09-19T09:35:16.721 & 1.97 & 1.403 & 26.49 \\ 
2020-09-19T09:35:17.811 & 1.90 & 1.403 & 26.50 \\ 
2020-09-19T09:35:18.901 & 1.83 & 1.403 & 26.52 \\ 
2020-09-19T09:35:19.991 & 1.76 & 1.403 & 26.53 \\ 
2020-09-19T09:35:21.081 & 1.69 & 1.403 & 26.54 \\ 
2020-09-19T09:35:22.171 & 1.62 & 1.403 & 26.56 \\ 
2020-09-19T09:35:23.261 & 1.55 & 1.403 & 26.57 \\ 
2020-09-19T09:35:24.351 & 1.48 & 1.403 & 26.58 \\ 
2020-09-19T09:35:25.441 & 1.41 & 1.403 & 26.60 \\ 
2020-09-19T09:35:26.531 & 1.34 & 1.403 & 26.61 \\ 
2020-09-19T09:35:27.622 & 1.27 & 1.403 & 26.62 \\ 
2020-09-19T09:35:30.892 & 1.07 & 1.403 & 26.66 \\ 
2020-09-19T09:35:31.437 & 1.03 & 1.403 & 26.67 \\ 
2020-09-19T09:35:31.982 & 1.00 & 1.403 & 26.67 \\ 
2020-09-19T09:35:32.527 & 0.96 & 1.403 & 26.68 \\ 
2020-09-19T09:35:33.072 & 0.93 & 1.403 & 26.69 \\ 
2020-09-19T09:35:33.617 & 0.90 & 1.403 & 26.69 \\ 
2020-09-19T09:35:34.162 & 0.86 & 1.403 & 26.70 \\ 
2020-09-19T09:35:34.707 & 0.83 & 1.403 & 26.71 \\ 
2020-09-19T09:35:35.252 & 0.79 & 1.403 & 26.72 \\ 
2020-09-19T09:35:35.797 & 0.76 & 1.403 & 26.72 \\ 
2020-09-19T09:35:36.342 & 0.73 & 1.403 & 26.73 \\ 
2020-09-19T09:35:36.887 & 0.69 & 1.403 & 26.74 \\ 
2020-09-19T09:35:37.432 & 0.66 & 1.403 & 26.74 \\ 
2020-09-19T09:35:37.977 & 0.62 & 1.403 & 26.75 \\ 
2020-09-19T09:35:38.522 & 0.59 & 1.403 & 26.76 \\ 
2020-09-19T09:35:39.067 & 0.56 & 1.403 & 26.76 \\ 
2020-09-19T09:35:39.612 & 0.53 & 1.403 & 26.77 \\ 
2020-09-19T09:35:40.157 & 0.49 & 1.403 & 26.78 \\ 
2020-09-19T09:35:40.702 & 0.46 & 1.403 & 26.78 \\ 
2020-09-19T09:35:41.247 & 0.42 & 1.403 & 26.79 \\ 
2020-09-19T09:35:41.792 & 0.39 & 1.403 & 26.80 \\ 
2020-09-19T09:35:42.337 & 0.36 & 1.403 & 26.80 \\ 
2020-09-19T09:35:42.882 & 0.33 & 1.403 & 26.81 \\ 
2020-09-19T09:35:43.427 & 0.30 & 1.403 & 26.82 \\ 
2020-09-19T09:35:43.972 & 0.27 & 1.403 & 26.82 \\ 
2020-09-19T09:35:44.517 & 0.24 & 1.403 & 26.83 \\ 
2020-09-19T09:35:45.062 & 0.22 & 1.403 & 26.84 \\ 
2020-09-19T09:35:45.607 & 0.19 & 1.403 & 26.84 \\ 
2020-09-19T09:35:46.152 & 0.17 & 1.403 & 26.85 \\ 
2020-09-19T09:35:46.697 & 0.16 & 1.403 & 26.86 \\ 
2020-09-19T09:35:47.242 & 0.15 & 1.403 & 26.86 \\ 
2020-09-19T09:35:47.787 & 0.15 & 1.403 & 26.87 \\ 
2020-09-19T09:35:48.332 & 0.15 & 1.403 & 26.88 \\ 
2020-09-19T09:35:48.877 & 0.16 & 1.403 & 26.89 \\ 
2020-09-19T09:35:49.422 & 0.17 & 1.403 & 26.89 \\ 
2020-09-19T09:35:49.967 & 0.19 & 1.403 & 26.90 \\ 
2020-09-19T09:35:50.512 & 0.22 & 1.403 & 26.91 \\ 
2020-09-19T09:35:51.057 & 0.24 & 1.403 & 26.91 \\ 
2020-09-19T09:35:51.602 & 0.27 & 1.403 & 26.92 \\ 
2020-09-19T09:35:52.147 & 0.29 & 1.403 & 26.93 \\ 
2020-09-19T09:35:52.692 & 0.33 & 1.403 & 26.93 \\ 
2020-09-19T09:35:53.237 & 0.36 & 1.403 & 26.94 \\ 
2020-09-19T09:35:53.782 & 0.39 & 1.403 & 26.95 \\ 
2020-09-19T09:35:54.327 & 0.42 & 1.403 & 26.96 \\ 
2020-09-19T09:35:54.872 & 0.45 & 1.403 & 26.96 \\ 
2020-09-19T09:35:55.417 & 0.48 & 1.403 & 26.97 \\ 
2020-09-19T09:35:55.962 & 0.52 & 1.403 & 26.98 \\ 
2020-09-19T09:35:56.507 & 0.55 & 1.403 & 26.99 \\ 
2020-09-19T09:35:57.052 & 0.58 & 1.403 & 26.99 \\ 
2020-09-19T09:35:57.597 & 0.62 & 1.403 & 27.00 \\ 
2020-09-19T09:35:58.143 & 0.65 & 1.403 & 27.01 \\ 
2020-09-19T09:35:58.688 & 0.68 & 1.403 & 27.01 \\ 
2020-09-19T09:35:59.233 & 0.72 & 1.403 & 27.02 \\ 
2020-09-19T09:35:59.778 & 0.75 & 1.403 & 27.03 \\ 
2020-09-19T09:36:00.323 & 0.78 & 1.403 & 27.03 \\ 
2020-09-19T09:36:00.868 & 0.82 & 1.403 & 27.04 \\ 
2020-09-19T09:36:01.413 & 0.85 & 1.403 & 27.05 \\ 
2020-09-19T09:36:01.958 & 0.88 & 1.403 & 27.06 \\ 
2020-09-19T09:36:02.503 & 0.92 & 1.403 & 27.06 \\ 
2020-09-19T09:36:03.048 & 0.95 & 1.403 & 27.07 \\ 
2020-09-19T09:36:03.593 & 0.98 & 1.403 & 27.08 \\ 
2020-09-19T09:36:05.773 & 1.11 & 1.403 & 27.11 \\ 
2020-09-19T09:36:07.408 & 1.21 & 1.403 & 27.13 \\ 
2020-09-19T09:36:09.043 & 1.31 & 1.403 & 27.15 \\ 
2020-09-19T09:36:10.678 & 1.41 & 1.403 & 27.17 \\ 
2020-09-19T09:36:12.313 & 1.51 & 1.403 & 27.19 \\ 
2020-09-19T09:36:13.948 & 1.61 & 1.403 & 27.22 \\ 
2020-09-19T09:36:15.583 & 1.71 & 1.403 & 27.24 \\ 
2020-09-19T09:36:17.218 & 1.81 & 1.403 & 27.26 \\ 
2020-09-19T09:36:18.853 & 1.91 & 1.403 & 27.28 \\ 
2020-09-19T09:36:20.488 & 2.01 & 1.403 & 27.31 \\ 
2020-09-19T09:36:22.123 & 2.10 & 1.403 & 27.33 \\ 
2020-09-19T09:36:23.758 & 2.20 & 1.403 & 27.35 \\ 
2020-09-19T09:36:25.393 & 2.30 & 1.403 & 27.38 \\ 
2020-09-19T09:36:27.028 & 2.40 & 1.403 & 27.40 \\ 
2020-09-19T09:36:28.664 & 2.50 & 1.403 & 27.42 \\ 
2020-09-19T09:36:30.299 & 2.60 & 1.403 & 27.44 \\ 
2020-09-19T09:36:31.934 & 2.69 & 1.403 & 27.47 \\ 
2020-09-19T09:36:33.569 & 2.79 & 1.403 & 27.49 \\ 
2020-09-19T09:36:35.204 & 2.89 & 1.403 & 27.51 \\ 
2020-09-19T09:36:36.839 & 2.98 & 1.403 & 27.54 \\ 
2020-09-26T02:18:23.003 & 1.51 & 1.407 & 31.49 \\ 
2020-09-26T02:18:24.093 & 1.45 & 1.407 & 31.51 \\ 
2020-09-26T02:18:25.183 & 1.39 & 1.407 & 31.52 \\ 
2020-09-26T02:18:26.273 & 1.34 & 1.407 & 31.54 \\ 
2020-09-26T02:18:27.363 & 1.28 & 1.407 & 31.56 \\ 
2020-09-26T02:18:28.453 & 1.22 & 1.407 & 31.58 \\ 
2020-09-26T02:18:29.543 & 1.17 & 1.407 & 31.59 \\ 
2020-09-26T02:18:30.633 & 1.11 & 1.407 & 31.61 \\ 
2020-09-26T02:18:31.723 & 1.05 & 1.407 & 31.63 \\ 
2020-09-26T02:18:32.813 & 1.00 & 1.407 & 31.64 \\ 
2020-09-26T02:18:33.903 & 0.94 & 1.407 & 31.66 \\ 
2020-09-26T02:18:34.993 & 0.89 & 1.407 & 31.68 \\ 
2020-09-26T02:18:36.083 & 0.84 & 1.407 & 31.70 \\ 
2020-09-26T02:18:37.173 & 0.78 & 1.407 & 31.71 \\ 
2020-09-26T02:18:38.263 & 0.73 & 1.407 & 31.73 \\ 
2020-09-26T02:18:39.354 & 0.68 & 1.407 & 31.75 \\ 
2020-09-26T02:18:40.444 & 0.62 & 1.407 & 31.77 \\ 
2020-09-26T02:18:41.534 & 0.57 & 1.407 & 31.78 \\ 
2020-09-26T02:18:43.714 & 0.48 & 1.407 & 31.82 \\ 
2020-09-26T02:18:44.259 & 0.46 & 1.407 & 31.83 \\ 
2020-09-26T02:18:44.804 & 0.43 & 1.407 & 31.84 \\ 
2020-09-26T02:18:45.349 & 0.41 & 1.407 & 31.84 \\ 
2020-09-26T02:18:45.894 & 0.40 & 1.407 & 31.85 \\ 
2020-09-26T02:18:46.439 & 0.38 & 1.407 & 31.86 \\ 
2020-09-26T02:18:46.984 & 0.36 & 1.407 & 31.87 \\ 
2020-09-26T02:18:47.529 & 0.34 & 1.407 & 31.88 \\ 
2020-09-26T02:18:48.074 & 0.33 & 1.407 & 31.89 \\ 
2020-09-26T02:18:48.619 & 0.32 & 1.407 & 31.90 \\ 
2020-09-26T02:18:49.164 & 0.30 & 1.407 & 31.91 \\ 
2020-09-26T02:18:49.709 & 0.30 & 1.407 & 31.92 \\ 
2020-09-26T02:18:50.254 & 0.29 & 1.407 & 31.92 \\ 
2020-09-26T02:18:50.799 & 0.29 & 1.407 & 31.93 \\ 
2020-09-26T02:18:51.344 & 0.29 & 1.407 & 31.94 \\ 
2020-09-26T02:18:51.889 & 0.29 & 1.407 & 31.95 \\ 
2020-09-26T02:18:52.434 & 0.30 & 1.407 & 31.96 \\ 
2020-09-26T02:18:52.979 & 0.30 & 1.407 & 31.97 \\ 
2020-09-26T02:18:53.524 & 0.31 & 1.407 & 31.98 \\ 
2020-09-26T02:18:54.069 & 0.33 & 1.407 & 31.99 \\ 
2020-09-26T02:18:54.614 & 0.34 & 1.407 & 32.00 \\ 
2020-09-26T02:18:55.159 & 0.36 & 1.407 & 32.01 \\ 
2020-09-26T02:18:55.704 & 0.37 & 1.407 & 32.01 \\ 
2020-09-26T02:18:56.249 & 0.39 & 1.407 & 32.02 \\ 
2020-09-26T02:18:56.794 & 0.41 & 1.407 & 32.03 \\ 
2020-09-26T02:18:57.339 & 0.43 & 1.407 & 32.04 \\ 
2020-09-26T02:18:57.884 & 0.46 & 1.407 & 32.05 \\ 
2020-09-26T02:18:58.429 & 0.48 & 1.407 & 32.06 \\ 
2020-09-26T02:18:58.974 & 0.50 & 1.407 & 32.07 \\ 
2020-09-26T02:18:59.519 & 0.52 & 1.407 & 32.08 \\ 
2020-09-26T02:19:00.064 & 0.55 & 1.407 & 32.09 \\ 
2020-09-26T02:19:01.699 & 0.62 & 1.407 & 32.11 \\ 
2020-09-26T02:19:03.334 & 0.70 & 1.407 & 32.14 \\ 
2020-09-26T02:19:04.969 & 0.78 & 1.407 & 32.17 \\ 
2020-09-26T02:19:06.604 & 0.85 & 1.407 & 32.19 \\ 
2020-09-26T02:19:08.239 & 0.93 & 1.407 & 32.22 \\ 
2020-09-26T02:19:09.875 & 1.01 & 1.407 & 32.25 \\ 
2020-09-26T02:19:11.510 & 1.09 & 1.407 & 32.28 \\ 
2020-09-26T02:19:13.145 & 1.17 & 1.407 & 32.31 \\ 
2020-09-26T02:19:14.780 & 1.25 & 1.407 & 32.33 \\ 
2020-09-26T02:19:16.415 & 1.34 & 1.407 & 32.36 \\ 
2020-09-26T02:19:18.050 & 1.42 & 1.407 & 32.39 \\ 
2020-09-26T02:19:19.685 & 1.50 & 1.407 & 32.42 \\ 
2020-09-26T02:19:21.320 & 1.58 & 1.407 & 32.44 \\ 
2020-09-26T02:19:22.955 & 1.66 & 1.407 & 32.47 \\ 
2020-09-26T02:19:24.590 & 1.74 & 1.407 & 32.50 \\ 
2020-09-26T02:19:26.225 & 1.82 & 1.407 & 32.53 \\ 
2020-09-26T02:19:27.860 & 1.90 & 1.407 & 32.56 \\ 
2020-09-26T02:19:29.495 & 1.99 & 1.407 & 32.59 \\ 
2020-09-26T02:19:31.130 & 2.07 & 1.407 & 32.62 \\ 
2020-09-26T02:19:32.765 & 2.15 & 1.407 & 32.64 \\ 
  \hline
\end{longtable}
\end{center}

\end{appendix}

\newpage

\end{document}